\documentstyle[aps,manuscript,epsf]{revtex}
\def\eff{_{{\rm eff}}}
\def\hpsi{{\hat\psi}}
\def\vg{{\vec g}}
\def\vr{{\vec r}}
\def\v0{{\vec 0}}
\def\hN{{\widehat N}}
\def\hn{{\hat n}}
\def\vnabla{{\vec \nabla}}

\def\vv{{\vec v}}

\def\vq{{\vec q}}
\def\hz{{\hat z}}
\def\hg{{\hat g}}

\def\myfigure#1#2{{\leftskip=0.10753\textwidth \rightskip\leftskip\small
\begin{figure}\baselineskip=14pt plus 2pt minus 1pt
\centerline{#1}\nobreak\smallskip\nobreak #2\end{figure}}}
\global\firstfigfalse

\title{Defects in Superfluids,  Superconductors and Membranes$^*$}
\author{David R. Nelson}
\address{Lyman Laboratory of Physics, Harvard University,
Cambridge, MA 02138}
\begin{document}
\draft
\maketitle
\widetext
\begin{abstract}
\baselineskip=16pt plus 2pt minus 1pt
An introduction to the defects which dominate the physics of superfluid
He$^4$ films, of superconducting slabs and of crystalline and hexatic
membranes is given. We first review point vortices in two-dimensional
neutral superfluids and discuss the unusual screening which arises when
the bosons are charged, as in superconducting films. Dislocation and
disclination defects in crystalline membranes are discussed from a
similar point of view. There is little or no screening in
``monolayer'' crystals, which are strongly constrained to lie in a
flat two-dimensional plane. A strong nonlinear screening effect
arises, however, in 2d membranes allowed to buckle into the third
dimension. This screening drastically lowers dislocation and
disclination energies, and forces crystalline membranes to melt at any finite
temperature. We point out that buckled 5- and 7-fold disclinations in
hexatic membranes have in general different logarithmically
divergent energies.
A similar asymmetry exists
in the energies of 5- and 7-fold defects in {\it liquid} membranes.
This difference determines the sign of the Gaussian bending rigidity,
and has important consequences in membranes which can change their topology or
with free boundary conditions.
\bigskip
\bigskip
\bigskip
\bigskip

\noindent
$^*$ Lectures given at the NATO ASI on Fluctuating Geometries in Statistical
Mechanics and Field Theory, Les Houches, Summer 1994.

\end{abstract}
\newpage
\baselineskip=19pt plus 2pt minus 1pt
\narrowtext
\section{INTRODUCTION}

Phase transitions and spontaneous symmetry breakings abound in modern condensed
matter physics. Of particular interest are the low energy excitations
associated with {\it continuous} broken symmetries, like the broken
translational
and orientational invariance of a crystal, the broken rotational symmetry
of ferromagnet or the broken
gauge symmetry of superfluids and superconductors. By imposing a slow spatial
variation of the continuous symmetry operation on the ground state, one
generates an important class of modes which can be easily excited near
$T=0$. In ferromagnets for example, these Goldstone modes are called spin
waves, or, in quantized form, ``magnons.'' In crystalline solids, the
translational
broken symmetryl leads to new shear phonon modes. Unlike a nematic liquid
crystal, the additional broken rotational symmetry of a solid does {\it not}
lead
to any extra low energy modes: Rotational modes are coupled to phonons in a
way reminiscent of the Higgs mechanism and are thus rendered massive---see
Ref. \cite{drn83} below and references therein. In superfluid helium, the
broken symmetry selects a particular phase for the condensate order parameter,
and leads to second sound. The physics of superconductors is complicated by the
{\it charge} of the Cooper pairs and by the coupling of electron motion to the
underlying ionic lattice. Low energy phase-degrees of freedom are still
important, although now it is the vector potential which acquires a mass
via the usual Higgs mechanism.

The excitations discussed above are all continuously connected to the ground
state.
However, point, line and wall defects in the order parameter texture, which
are {\it not} continuously connected to the ground state, may also be
important. Dislocation lines mediate plastic flow, and point-like vacancies and
interstitials allow rapid particle diffusion in crystalline solids.
Vortex rings control the decay of supercurrents in superfluids and
superconductors. Most laboratory ferromagnets are subject to weak crystal
fields, which break rotational invariance. In this case, the width of
hysterisis loops below the Curie temperature is controlled by the motion of
bloch walls. Under some circumstances, such topological defects also play an
important role in the phase transition out of the broken symmetry state.

In this review, we discuss the defects which arise in superfluids,
superconductors
and membranes. Although the emphasis is how to incorporate defects into simple
Landau-Ginzburg field theories, we also touch upon their role in the
statistical mechanics. We concentrate on two-dimensional systems, so that the
defects are typically point-like.

The classic example of vortices and the Kosterlitz-Thouless transition in
superfluid helium films is discussed first, together with universal jump in
the superfluid density. Experimental confirmation of this prediction was
an important test of the idea of defect-mediated phase transitions.
{\it Screening} of defect energies in superconducting films is then treated,
with
particular attention to the spreading of magnetic field lines in the
three-dimensional volume outside the sample. When this spreading is
ignored, screening efficiently cuts off the logarithmic divergence in the
vortex energy. The finite energy vortices in this ``naive'' theory
then proliferate for entropic reasons at any finite temperature, thus
destroying superconductivity. In realistic samples, however, magnetic
field line spreading (which arises because supercurrents are confined to a
two-dimensional plane) renders screening less efficient, and a
Kosterlitz-Thouless transition from a finite temperature superconducting
state becomes possible.

The remainder of the paper contrasts the behavior of defects in membranes and
monolayers. By ``monolayers,'' we mean crystalline films strongly confined
to a plane. Dislocations and disclinations control the statistical
mechanics of monolayers in a way similar to vortices in neutral
superfluids. Successive dislocation and disclination unbinding
transitions lead from crystalline to hexatic to liquid monolayer
phases. In contrast to monolayers, ``membranes'' have a relatively low
energy cost to buckle out of the two-dimensional plane, thus escaping
into the third dimension. As a result, defects can be screened by
out-of-plane displacements much like vortices in the ``naive'' model of
superconducting films. Dislocations, for example, now have a finite
energy. Their entropic proliferation causes crystals to be replaced by
{\it hexatics} as the inevitable low temperature phase of membranes. We
also discuss the buckling of 5- and 7-fold disclinations, of vacancies,
interstitial and impurity atoms, and of grain boundaries.

We conclude by emphasizing the fundamental asymmetry between 5- and 7-fold
disclinations in membranes. There is an exact symmetry between plus and
minus vortices in superfluids and between dislocations with equal and opposite
burgers vectors in two-dimensional crystals, which forces the energies of the
defect and anti-defects to be equal. No such symmetry, however, insures
that the core energies of plus and minus disclinations are identical.
These disclinations represent points of local 5- and 7-fold coordination,
where the number of neighboring particles is determined by the Dirichlet
construction. This situation is similar to vacancies and interstitials in
crystalline solids: Vacancies typically have a lower energy than
interstitials, even though these excitations are ``anti-defects'' of
each other.

Disclination asymmetry is of little consequence in flat monolayers, where
the concentrations of 5- and 7-fold defects must be equal for topological
reasons. (More precisely, Euler's theorem ensures that the average
coordination is six in a sufficiently large system.) In hexatic monolayers,
moreover, the logarithmic divergences in the energies of isolated plus
and minus disclinations have equal coefficients. The situation is
different, however, in {\it membranes} which are free to change their
topology or with free boundary conditions which allow them to curl up near
the edges. When the bending rigidity which controls out-of-plane fluctuations
is small, buckling in hexatic membranes leads to different coefficients in
the logarithmically diverging disclination energies. In liquid membranes,
moreover, different core energies of plus and minus disclinations will
inevitably bias the system towards a net positive or negative average
Gaussian curvature. The energy asymmetry between 5- and 7-fold
disclinations in the liquid determines the sign of the Gaussian
rigidity, which acts as chemical potential controlling their
population difference. Returning to the comparison with a superconductor,
the Gaussian rigidity acts like an external magnetic field, which would bias
superconducting films toward a net excess of plus or minus vortices. In
membranes, the sign of the Gaussian bending rigidity determines whether
vesicle phases (with positive net Gaussian curvature) or ``plumbers
nightmare'' phases (with overall negative Gaussian curvature) are preferred.

Theories and simulations of hexatic and liquid membranes usually avoid
this issue by imposing a definite spherical or torroidal topology, which
automatically fixes the difference in the number of 5- and 7-fold defects.
Such constraints are clearly artificial, however, for lipid bilayer
membranes which can change their topology at will, and may even have free edges
at some intermediate stage of a topological change. The analogous constraint
for vacancies and interstitials in solids would be imposed by periodic
boundary conditions: these defects can then only be created in pairs. In
real solids, with free boundary conditions, vacancies and interstitials
can enter at the surface to produce the concentrations dictated by the
microscopic energetics. The simplest probe of disclination bias would be to
simulate liquid membranes with free edges, and see which defect begins to
dominate as the bending rigidity is reduced from large values.
See section~III.D for details.

\section{TWO-DIMENSIONAL SUPERFLUIDS AND SUPERCONDUCTORS}
\label{sec:two-dimen}

A detailed and successful theory of two-dimensional superfluidity was
constructed in the 1970s. The theory explains the observed properties of thin
films of superfluid He$^4$ on various substrates and the physics of thin
superconducting films. The basic physical idea of defect unbinding was
proposed by Kosterlitz and Thouless \cite{kosthou72}, by Berezinski
\cite{bere70},
and, in a different physical context, by deGennes \cite{degen}. The first
detailed calculations were carried out by Kosterlitz \cite{kos74}, who
exploited a pioneering renormalization group method developed by Anderson and
Yuval for the Kondo problem \cite{andyu}. Because extensive reviews are
available elsewhere \cite{kosthou78,bihrev,drn83}, we limit ourselves to a
brief sketch of the models and the basic results. We shall see how vortices
control the basic physics and discuss their statistical mechanics. We discuss
only equilibrium properties, and ignore the important area of superfluid
and superconducting dynamics \cite{ambetal,bihdrn79}.

\myfigure{\epsfysize3in\epsfbox{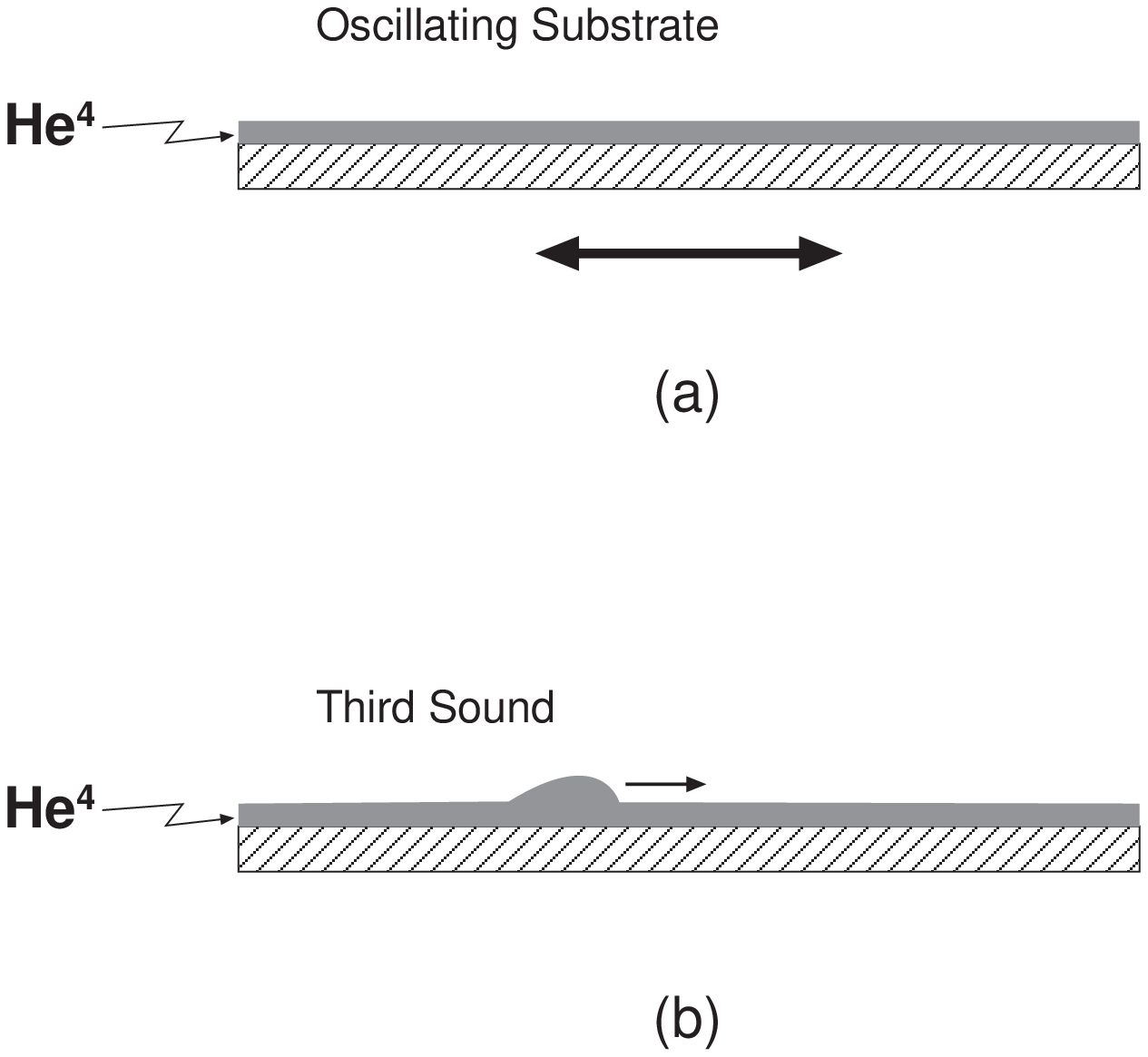}}{FIG.~1. \
(a) Schematic of an oscillating substrate experiment, which measures
the superfluid density.  (b) Propagation of a
third sound excitation. Measuring the dispersion relation also allows one
to extract $\rho_s(T)$.}

\subsection{Superfluid Helium Films: Experimental Facts}
\label{subsec:superfluid}

Superfluidity arises with decreasing temperature in helium films in a
fundamentally different way than in bulk materials. Two basic experiments
which reveal this difference are illustrated in Fig.~1. On an
oscillating substrate, a finite fraction of the He$^4$ film decouples from the
underlying motion below a critical temperature $T_c$. The film appears
lighter than it should be, and measuring this deficit allows one to extract
the temperature-dependent superfluid density $\rho_s(T)$ \cite{bis78}. In a
related
experiment, third-sound excitations similar to capillary waves propagate
below $T_c$ with negligible damping through films as thin as
a fraction of an Angstrom \cite{rud78}.
Such propagating waves would be impossible in a classical fluid, due to viscous
damping. The primary restoring force is the Van der Waals attraction to the
substrate
(rather than gravity or surface tension), and again one can extract
$\rho_s(T)$ provided the temperature is not too close to $T_c$, where the
waves become strongly damped.

\myfigure{\epsfysize2.5in\epsfbox{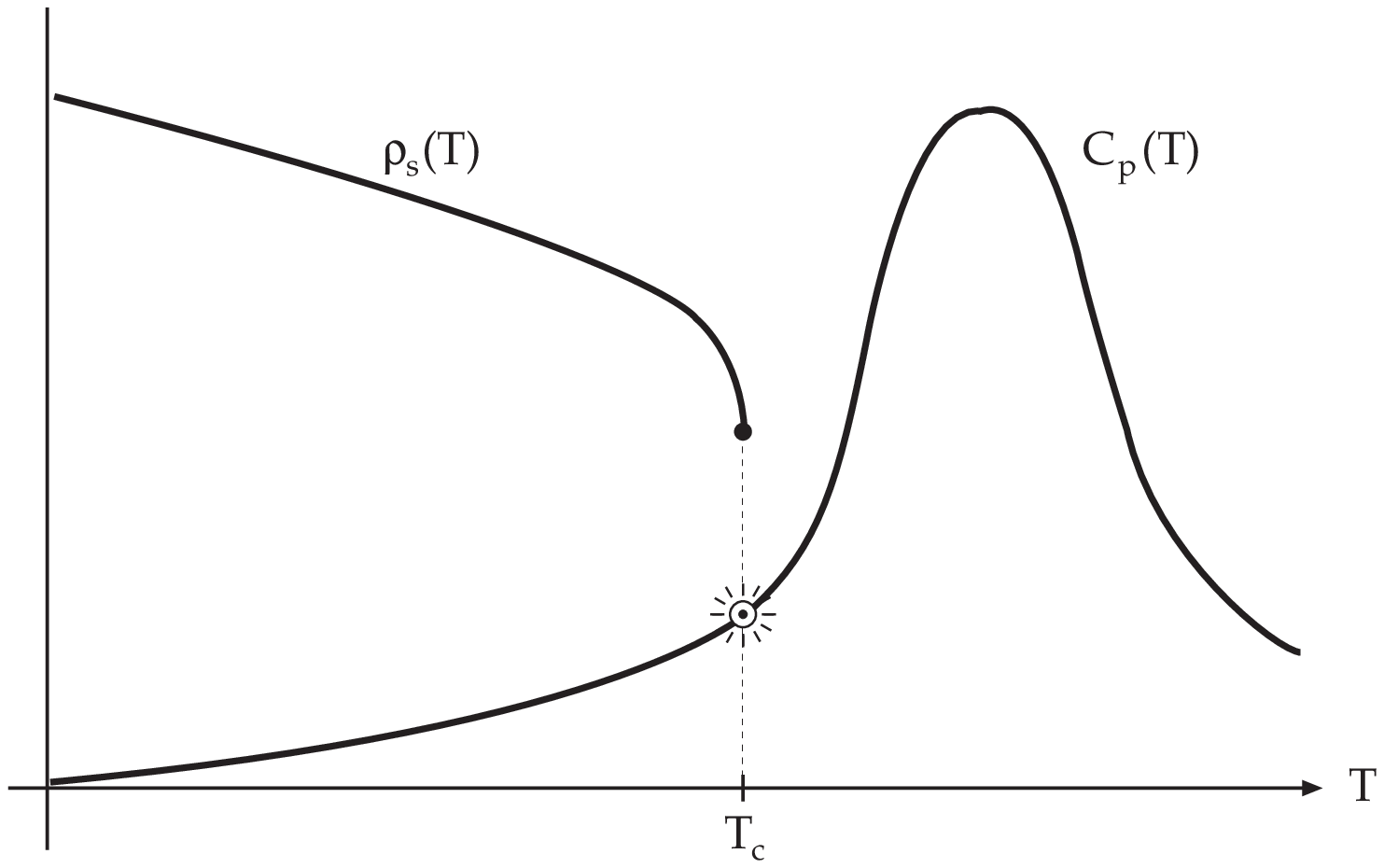}}{FIG.~2. \
 Jump discontinuity in the superfluid density $\rho_s(T)$ as a
function of temperature for a fixed thickness helium film. Also shown is the
superfluid specific heat at constant pressure
as a function of temperature. There is only an
essential singularity, $C_p(T)\approx c_1+c_2\exp [-c_3/|T-T_c|^{1/2}]$, near
the transition, where the $c_i$ are constants.}

As illustrated in Fig.~2, the superfluid density in films actually
jumps discontinuously to zero at $T_c$, in contrast to the continuous vanishing
observed in bulk superfluid He$^4$. The phase transition is nevertheless
{\it not} first
order,  as becomes evident from an examination of the specific heat
(see Fig.~3). There is only a rounded maximum {\it above} the
transition temperature, in contrast to the delta function peak
expected for a true first-order transition. The detailed theory shows that
the specific heat contains only an unobservable essential singularity at the
transition itself \cite{gas,berdrn79}.

\subsection{Theoretical Background}
\label{subsec:theoback}

Understanding the equilibrium behavior of superfluid helium films at
temperature
$T$ requires calculation of quantities such as the quantum partition function,
\begin{equation}
Z=Tr'\{e^{-\widehat{\cal H}/T}\} \; ,
\eqnum{2.1}
\label{eq:one}
\end{equation}
where $\cal H$ is the Hamiltonian operator,
\begin{equation}
\widehat{\cal H}=
{-\hbar^2\over 2m}\;
\sum_{j=1}^N\;
\nabla_j^2+\sum_{i>j}\;
V(|\vec r_i-\vec r_j|) \; ,
\eqnum{2.2}
\label{eq:two}
\end{equation}
and $Tr'$ means a sum over all {\it symmetrized} energy eigenfunctions,
reflecting the bosonic nature of a He$^4$ atom. Here $V(r)$ is the boson pair
potential, and we have set Boltzmann's constant $k_B=1$. Explicit computation
of
$Z$ for a fixed large number of particles $N$ by finding all symmetric energy
eigenfunctions (or even just the ground-state wave function \cite{mah}) is
difficult
and it is useful to bypass this problem by changing to a second
quantized representation of the many body physics. In second quantized
notation, the Hamiltonian (\ref{eq:two}) reads
\begin{eqnarray}
\widehat{\cal H} &=& \int d^3r\;\hpsi^+
(\vr)\left({-\hbar^2\over 2m}\;\nabla^2\right)
\hpsi(\vr)\nonumber\\
&\quad +&{1\over 2}
\int d^3r\int d^3r'\;\hpsi^+
(\vr)\hpsi^+(\vr\;')
V(|\vr-\vr\;'|)
\hpsi(\vr\;')\hpsi(\vr)\; ,
\eqnum{2.3}
\label{eq:three}
\end{eqnarray}
where $\hpsi^+(\vr)$ and $\psi(\vr)$ are boson creation and
destruction operators obeying the commutation
relations,
\begin{equation}
[\hpsi(\vr),\hpsi^+(\vr\;')]=
\delta(\vr-\vr\;'),
\eqnum{2.4}
\label{eq:four}
\end{equation}
with all other commutators zero. Within the second quantization formalism,
it is convenient to calculate  thermodynamic quantities via the {\it grand}
canonical partition function,
\begin{equation}
Z_{gr}=Tr'\left\{e^{-(\widehat{\cal H}-\mu\widehat N)/T}\right\},
\eqnum{2.5}
\label{eq:five}
\end{equation}
where $\mu$ is the chemical potential and the number operator is
\begin{equation}
\hN=\int d^3r\;\hpsi^+(\vr)\hpsi(r)\; .
\eqnum{2.6}
\label{eq:six}
\end{equation}
As pointed out by Penrose and Onsager \cite{penons}, Bose condensation in
bulk He$^4$ is associated with off--diagonal long-range order in the
correlation function,
\begin{equation}
G(r)=\langle\psi^+(\vr)\psi(\v0)\rangle,
\eqnum{2.7}
\label{eq:seven}
\end{equation}
where $\langle Q\rangle$ means $Tr'\{Q\exp[-(\widehat{\cal
H}-\mu\hN)/T]\}/Z_{gr}$.
Specifically, one has
\begin{eqnarray}
\lim_{r\rightarrow\infty} G(\vr)&=&
\langle\hpsi^+(\vr)\rangle\;\langle\hpsi(\v0)\rangle \nonumber \\
&\not=& 0\;.
\eqnum{2.8}
\label{eq:eight}
\end{eqnarray}

A simplified description of superfluidity in terms of a {\it Landau} theory
results from a coarse graining procedure. In the case of helium films,
we start by defining a $c$-number field
\begin{equation}
\psi(\vr)=\langle\hpsi\rangle_{\Omega(\vr)},
\eqnum{2.9}
\label{eq:nine}
\end{equation}
where the brackets include a spatial average over a volume $\Omega(\vr)$
centered on $\vr$ which is large compared to the spacing between helium
atoms but small compared to the overall system size. The coarse-grained
fields $\psi^*(\vr)$ and $\psi(\vr)$ (which are nonzero when the He$^4$
liquid is superfluid in view of Eq.\ (\ref{eq:eight})) behave like complex
numbers (instead of operators) for large enough averaging volumes. In
helium films, we also take the averaging size to be large compared to the
film thickness. The long wavelength spatial configurations of the system
are now specified by complexions of $\psi(\vr)$, where $\vr$ is a
{\it two}-dimensional spatial variable, and the partition function is
a functional integral,
\begin{equation}
Z_{gr}=\int\;{\cal D} \psi(\vr)
\exp[-F/T]\;.
\eqnum{2.10}
\label{eq:ten}
\end{equation}
The free energy $F$ includes entropic contributions due to the coarse-graining
procedure. The Landau expansion of $F/T$ in the order parameter $\psi(r)$
has the same form as $\widehat{\cal H}-\mu\hN$, with $\hpsi(\vr)$ and
$\hpsi^+(\vr)$
replaced by the $c$-number fields $\psi(\vr)$ and $\psi^*(\vr)$,
\begin{equation}
{F\over T}=
\int\; d^2r\left[{1\over 2}\;A|\vnabla\psi|^2+
{1\over 2}\;a|\psi|^2+b|\psi|^4+
\cdots\right]
\eqnum{2.11}
\label{eq:eleven}
\end{equation}
where, as usual, $a$ changes sign at the mean-field transition temperature
$T_c^0$, $a=a'(T-T_c^0)$.

Because the long wavelength fluctuations in a helium film are two-dimensional,
the actual temperature $T_c$ where true long-range
order develops is suppressed well below
$T_c^0$. For $T\approx T_c$, $a\ll 0$ and the polynomial part of $F$
has a deep minimum, even though thermal fluctuations may nevertheless suppress
genuine diagonal long-range order at the largest length scales. To a
first approximation, we can then neglect fluctuations in the {\it amplitude}
of $\psi(r)$, and set
\begin{equation}
\psi(\vr)\approx\psi_0e^{i\theta(\vr)},
\eqnum{2.12}
\label{eq:twelve}
\end{equation}
where $\psi_0\approx\sqrt{-a/4b}$ and $\theta(\vr)$ is a slowly varying phase
variable. The free energy becomes
\begin{equation}
{F\over T}=
{\rm const.} +
{1\over 2}
\;K_0\int |\vnabla\theta(\vr)|^2\;d^2r,
\eqnum{2.13}
\label{eq:thirteen}
\end{equation}
where $K_0=A\psi_0^2$. The physical interpretation of $K_0$ becomes clear if we
recall
that the superfluid velocity $\vv_s$ is given by \cite{khal}
\begin{equation}
\vv_s(\vr)={\hbar\over m}\;
\vnabla\theta(\vr),
\eqnum{2.14}
\label{eq:fourteen}
\end{equation}
and write the contribution to $F$ of the suprfluid kinetic energy in
terms of the superfluid density $\rho_s^0$ as
\begin{equation}
F={\rm const.} + {1\over 2} \rho_s^0\int\;d^2r
|\vv_s(\vr)|^2\;.
\eqnum{2.15}
\label{eq:fifteen}
\end{equation}
Evidently, $K_0$ is related to the superfluid density in this
``phase only'' approximation by
\begin{equation}
K_0={\hbar^2\over m^2T}\;\rho_s^0\;.
\eqnum{2.16}
\label{eq:sixteen}
\end{equation}

The peculiar nature of superfluid order in two dimensions becomes evident if
we now evaluate $G(r)$, by integrating freely over the phase
field when evaluating averages weighted by
$e^{-F/T}$,
\begin{eqnarray}
G(r)&=&\langle\psi(\vr)\psi^*(\v0)\rangle\nonumber \\
&=& \psi_0^2\exp\left[-{1\over 2}\langle[\theta(\vr)-
\theta(\v0)]^2\rangle\right]\;,
\eqnum{2.17}
\label{eq:seventeen}
\end{eqnarray}
where the last line is a general property of Gaussian fluctuations.
Upon introducing Fourier variables, the remaining average is easily
evaluated using the equipartition theorem,
\begin{eqnarray}
\langle[\theta(\vr)-\theta(\v0)]^2\rangle
&=& {2\over K_0}\int{d^2q\over (2\pi)^2}
\;{1\over q^2}(1-e^{i\vq\cdot\vr)}\nonumber \\
&\mathop{\approx}\limits_{r\rightarrow \infty}&{1\over  K_0}
\ln(r/a_0)\;,
\eqnum{2.18}
\label{eq:eighteen}
\end{eqnarray}
where $a_0$ is a microscopic cutoff. We thus find that $G(r)$
decays {\it algebraically} to zero \cite{weg},
\begin{equation}
G(r)\sim 1/r^{\eta(T)}\;,
\eqnum{2.19}
\label{eq:nineteen}
\end{equation}
with
\begin{equation}
\eta(T)=1/2\pi K_0(T)\;.
\eqnum{2.20}
\label{eq:twenty}
\end{equation}
Thus there is only {\it quasi}-long range off-diagonal long-range order at any
finite temperature in two dimensions. Nevertheless, this algebraic order is
enough to distinguish the low-temperature superfluid from a distinct high
temperature normal liquid film in which $G(r)$ decays exponentially,
\begin{equation}
G(r)\sim\exp[-r/\xi(T)]\;,
\eqnum{2.21}
\label{eq:twentyone}
\end{equation}
which defines a superfluid coherence length $\xi(T)$. Note that the
low-temperature phase still has a nonzero superfluidity density, similar to
bulk superfluids. In contrast to bulk superfluids, however,  the
low-temperature
phase is characterized by a {\it continuously varying} critical exponent
$\eta(T)$.

\myfigure{\epsfysize5in\epsfbox{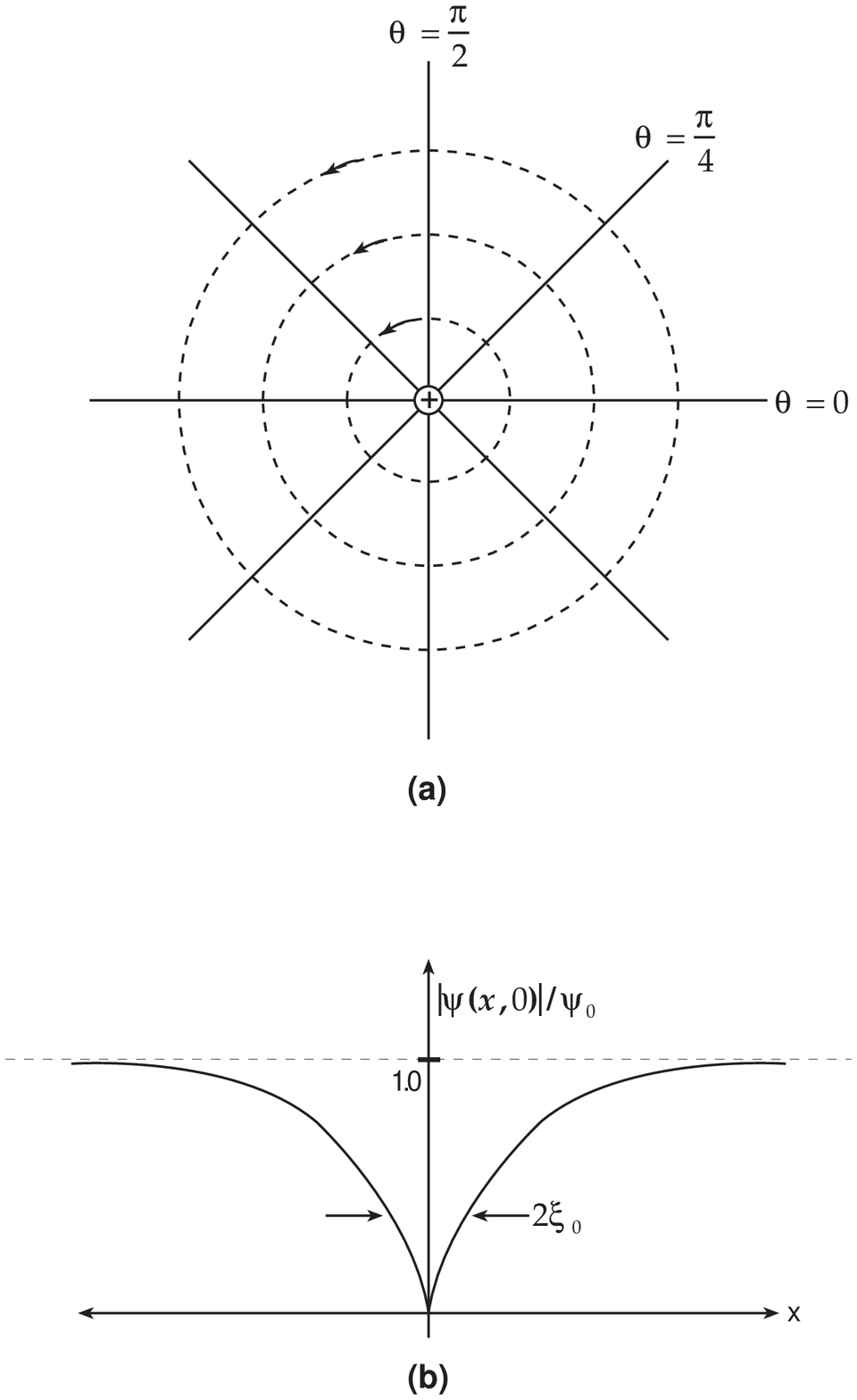}}{FIG.~3. \
(a) Lines of constant phase associated with a vortex in a
two-dimensional superfluid. Dashed circles are the streamlines for the
associated supercurrents. (b) Variation of the order parameter magnitude near a
vortex core located at the origin.}

The ``phase only'' description used above neglects amplitude fluctuations,
and ignores the periodic nature of the phase variable. We can account for both
effects by introducing a discrete set of vortex singularities in $\psi(\vr)$.
As illustrated in Fig.~3a, vortices represent zeroes of $\psi(\vr)$, where the
phase is undetermined. Near a vortex core, the assumption that the order
parameter amplitude $|\psi(\vr)|$ is a constant breaks down. Fig.~3b
shows how $\psi(\vr)$ varies in the vicinity of a vortex \cite{kawa}.
$\psi(\vr)$ rises from zero to its bulk value over a characteristic
length $\xi_0$, related to  the coefficients of the first two terms
of Eq. (\ref{eq:eleven}),
\begin{equation}
\xi_0=\sqrt{A/|a|}\;.
\eqnum{2.22}
\label{eq:twentytwo}
\end{equation}

To describe a radially symmetric net phase change of $\pm 2\pi$
near a vortex centered at the origin, we set
\begin{equation}
\theta(x,y)=\pm\tan^{-1}(y/x)\;.
\eqnum{2.23}
\label{eq:twentythree}
\end{equation}
According to Eq. (\ref{eq:fourteen}) we then have a velocity field
\begin{equation}
\vv_s={\hbar\over m}\;
{\hz\times\vr\over r^2}\;,
\eqnum{2.24}
\label{eq:twentyfour}
\end{equation}
which leads to a vortex energy
\begin{equation}
E_v=\pi\rho_s^0{\hbar^2\over m^2}\;\ln(R/\xi_0)+C
\eqnum{2.25}
\label{eq:twentyfive}
\end{equation}
when inserted into Eq. (\ref{eq:fifteen}), where, $R$ is the system size.
We have imposed a lower cutoff $\xi_0$ and included the contribution from
smaller scales where the amplitude varies in the constant $C$. Since there are
approximately $(R/\xi_0)^2$ distinct positions for a vortex, the {\it free}
energy $E_v-TS_v$ for large $R$ is
\begin{equation}
F_v\approx\left(\pi\rho_s^0{\hbar^2\over m^2}-
2T\right)\ln(R/\xi_0)\;.
\eqnum{2.26}
\label{eq:twentysix}
\end{equation}
The prediction of this famous argument \cite{kosthou72,bere70,degen}
is that it becomes favorable for vortices to proliferate (leading to
exponential decay of $G(r)$) above a critical temperature
\begin{equation}
T_c\approx{\pi\over 2}{\hbar^2\over m^2}\;\rho_s^0\;.
\eqnum{2.27}
\label{eq:twentyseven}
\end{equation}

\subsection{Vortex Statistical Mechanics and Renormalization of $\rho_s$}
\label{subsec:statistical}

Going beyond the simple argument presented above requires a more detailed
theory, which allows for a finite density of interacting vortices near
$T_c$. Understanding superfluidity in two-dimensional helium films is
equivalent to solving the statistical mechanics of this vortex gas. We
start by noticing that functional integrals like the partition function
(\ref{eq:ten}) are dominated in the low-temperature ``phase only''
approximation by solutions of
\begin{equation}
\nabla^2\theta=0\;.
\eqnum{2.28}
\label{eq:twentyeight}
\end{equation}
To allow for vortices (and thus for amplitude fluctuations), we require
that Eq. (\ref{eq:twentyeight}) be satisfied ``almost everywhere,'' i.e.,
everywhere except in the cores of a collection $N$ of vortices located
at positions $\{\vr_\alpha\}$ with integer charges $\{s_\alpha\}$. A vortex
singularity
has ``charge'' $s_\alpha$ if the line integral of the phase gradient on any
path
enclosing the core satisfies,
\begin{equation}
\oint\;\vnabla\theta\cdot d\vec\ell=
2\pi s_\alpha\;.
\eqnum{2.29}
\label{twentynine}
\end{equation}
The function (\ref{eq:twentythree}) obeys this condition, for
example, with $s_\alpha=\pm 1$. More generally we expect, for a contour $\cal
C$
enclosing many vortices, that
\begin{equation}
\oint_{\cal C}\vnabla\theta\cdot d\ell=
\int_\Omega d^2r\;n_v(\vr)\;,
\eqnum{2.30}
\label{eq:thirty}
\end{equation}
where $\Omega$ is the area spanned by $\cal C$ and the vortex ``charge
density''
is
\begin{equation}
n_v(\vr)=2\pi\sum_{\alpha=1}^N\;
s_\alpha\delta(\vr-\vr_\alpha)\;.
\eqnum{2.31}
\label{eq:thirtyone}
\end{equation}

Equation (\ref{eq:thirty}) is a statement about the noncommutivity of
derivatives of $\theta(x,y)$. Indeed, by taking the contour $\cal C$ to be a
small square loop, we readily find
\begin{eqnarray}
\epsilon_{ij}\partial_i\partial_j\theta(\vr)&=&
\partial_x\partial_y\theta-\partial_y\partial_x\theta\nonumber \\
&=& n_v(\vr)\;,
\eqnum{2.32}
\label{eq:thirtytwo}
\end{eqnarray}
where $\epsilon_{ij}$ is the antisymmetric unit tensor in two
dimensions, $\epsilon_{xy}=-\epsilon_{yx}=1$. To cast this equation in a
more familiar form, we introduce the Cauchy conjugate to the phase field,
by
\begin{equation}
\partial_i\theta(\vr)=\epsilon_{ij}\partial_j
\tilde\theta(\vr)\;,
\eqnum{2.33}
\label{eq:thirtythree}
\end{equation}
and find that $\tilde\theta(x,y)$ satisfies
\begin{equation}
\nabla^2\tilde\theta(\vr)=n_v(\vr)\;.
\eqnum{2.34}
\label{eq:thirtyfour}
\end{equation}
Finding the phase associated with a set of vortex singularities thus requires
that we first determine the ``electrostatic potential'' $\tilde\theta(\vr)$
of a collection of point charges $\{s_j\}$ at positions $\{\vr_j\}$ in two
dimensions \cite{kosthou72}. The solution of Eq. (\ref{eq:thirtyfour}) is
\begin{equation}
\tilde\theta(\vr)=2\pi\sum_{\alpha}\;s_\alpha G(\vr,\vr_\alpha)\;,
\eqnum{2.35}
\label{eq:thirtyfive}
\end{equation}
where the Green's function satisfies
\begin{equation}
\nabla^2G(\vr,\vr_\alpha)=\delta(\vr-\vr_\alpha)\;.
\eqnum{2.36}
\label{eq:thirtysix}
\end{equation}
For $|\vr-\vr_j|$ large and both points far from any boundaries, we have
\begin{equation}
G(\vr,\vr_j)\approx{1\over 2\pi}\ln\left(
{|\vr-\vr_j|\over
\xi_0}\right)+C\;,
\eqnum{2.37}
\label{eq:thirtyseven}
\end{equation}
where $C$ is a constant which contributes to the vortex core energy.

We now decompose the phase into a contribution $\theta_v(x,y)$ from vortices,
obtained by taking the Cauchy conjugate of Eq. (\ref{eq:thirtyfive}),
and a smoothly varying part $\phi(x,y)$,
\begin{equation}
\theta(x,y)=\theta_v(x,y)+\phi(x,y)\; .
\eqnum{2.38}
\label{eq:thirtyeight}
\end{equation}
The function $\phi(x,y)$ represents single-valued phase fluctuations
superimposed on the vortex extrema. Thermodynamic averages are obtained
by first integrating over this nonsingular phase field, and then
summing over all possible complexions of vortex charges and positions.
To insert this decomposition into Eq. (\ref{eq:thirteen}), we need
\begin{equation}
\vec \nabla\theta(\vr)=2\pi(\hz\times\vnabla)\int d^2r'
n(\vr\;')G(\vr,\vr\;')+\vnabla\phi\;.
\eqnum{2.39}
\label{eq:thirtynine}
\end{equation}
The resulting free energy takes the form \cite{kosthou72}
\begin{equation}
{F\over T}={\rm const.} +
{1\over 2} K_0\int d^2r|\vnabla\phi|^2+
{F_v\over T}\;,
\eqnum{2.40}
\label{eq:forty}
\end{equation}
where the vortex part is
\begin{equation}
{F_v\over T}
=-\pi K_0\sum_{\alpha\not=\beta}
\;s_\alpha s_\beta\ln\;
\left({|\vr_\alpha-\vr_\beta'|\over a}\right)
+{E_c\over T}\sum_\alpha s_\alpha^2\;,
\eqnum{2.41}
\label{eq:fortyone}
\end{equation}
and the core energy $E_c$ is usually assumed to be proportional to
$K_0$. Implicit in the statistical mechanics associated with
Eq. (\ref{eq:fortyone}) is a constraint of overall
``charge neutrality,'' $\sum_j s_j=0$, required for a finite
energy in the thermodynamic limit.

\subsection{Renormalization Group and Universal Jump in the
Superfluid Density}
\label{subsec:renormal}

To illustrate the statistical mechanics of the vortex gas described
above, consider the renormalized superfluid density
$\rho_s^R(T)$ calculated to lowest order in the vortex
fugacity
\begin{equation}
y=e^{-E_c/T}\;.
\eqnum{2.42}
\label{eq:fortytwo}
\end{equation}
The renormalized superfluid density is related to the correlations of
the momentum density $\vec g(\vr)$.
On a microscopic level, the  momentum density
operator $\hg_i(\vr)$ is given in terms particle creation and
destruction operators by
\begin{equation}
\hg_i(\vr)={\hbar\over 2i}
\left[\hpsi^+(\vr)\partial_i\hpsi(\vr)-
\hpsi(\vr)\partial_i\hpsi^+(\vr)\right]
\;.
\eqnum{2.43{\rm a}}
\label{eq:fortythreea}
\end{equation}
When local off-diagonal long-range order is present in helium films,
we replace $\hpsi^+(\vr)$ and $\hpsi(\vr)$ by coarse-grained
two-dimensional classical $c$-number fields, as usual.
In the ``phase only'' approximation which led to
Eq. (\ref{eq:thirteen}), the coarse-grained momentum density is then
\begin{equation}
\vg(\vr)=\rho_s^0\vv_s(\vr)\;,
\eqnum{2.43{\rm b}}
\label{eq:fortythreeb}
\end{equation}
where $\rho_s^0=m|\psi_0|^2$, and $\vv_s(\vr)$ includes possible
contributions from vortices. In helium films, the contribution of
the {\it normal} fluid to the momentum vanishes, due to the
viscous coupling to the substrate.

The correlation matrix which determines the renormalized
superfluid density is
\begin{equation}
C_{ij}(\vq,K,y)\equiv
\langle g_i(\vq)g_j^*(\vq)\rangle\;,
\eqnum{2.44{\rm a}}
\label{eq:fortyfoura}
\end{equation}
where $g_i(\vq)$ is the Fourier transform of the $i$-th
component of $\vg(\vr)$. To extract the renormalized
superfluid density, we first decompose $C_{ij}$ into transverse
and longitudinal parts:
\begin{equation}
C_{ij}=A(q)
{q_iq_j\over q^2}
+B(q)\left(
\delta_{ij}-{q_iq_j\over q^2}\right)\;.
\eqnum{2.44{\rm b}}
\label{eq:fortyfourb}
\end{equation}
In an isotropic classical liquid, one would have $A(q)=B(q)$ in the limit
$q\rightarrow 0$. The momentum fluctuations, moreover, would
decouple from the configurational degrees of freedom responsible
for most phase transitions. The behavior of quantum fluids is
different: The renormalized superfluid density $\rho_s^R(T)$,
in particular, is given by the difference between $A$ and $B$
as $q$ tends to zero (see Appendix)
\begin{equation}
\rho_s^R(T)={1\over T}\lim_{q\rightarrow 0}\;
[A(q)-B(q)]\;.
\eqnum{2.45}
\label{eq:fortyfive}
\end{equation}

The momentum density $\vg(\vr)=\rho_s{\hbar\over m}
\vnabla\theta(\vr)$ is already decomposed into transverse and
longitudinal  parts in Eq. (\ref{eq:thirtynine}).
One readily finds that $C_{ij}$ takes the form
(\ref{eq:fortyfourb}), with \cite{drn83}
\begin{eqnarray}
{\hbar^2\over m^2T}\; A(q)&=& K_0\eqnum{2.46{\rm a}}
\label{eq:fortysixa} \\
{\hbar^2\over m^2T}B(q)&=&
{4\pi^2K_0^2\over q^2}
\langle \hn_v(\vq)\hn_v(-\vq)\rangle\;.
\eqnum{2.46{\rm b}}
\label{eq:fortysixb}
\end{eqnarray}
where $\hn_v(\vq)$ is the Fourier transform of the vortex density
$n_v(\vec r)$, and the average in (\ref{eq:fortysixb}) is to be
carried out over the vortex part of the free energy.

When the vortices are dilute, it is straightforward to use these
results to obtain a fugacity perturbation expansion for the
renormalized superfluid density. In terms of $K_R\equiv
\hbar^2\rho_s^R/m^2T$, we have \cite{drn83}
\begin{equation}K_R^{-1}=K_0^{-1}
+4\pi^3y_0^2\int_a^\infty
{dr\over a}\left({r\over a}\right)
^{3-2\pi K_0}+
O(y_0^4)\;.
\eqnum{2.47}
\label{eq:fortyseven}
\end{equation}
At low temperatures, Eq. (\ref{eq:fortyseven}) provides a
small correction (proportional to $e^{-2E_c/T}$) to $K_R^{-1}$. When
$K_0\;\alt\;
2/\pi$, however, the integral becomes infrared divergent and
perturbation theory breaks down. This is precisely the condition
(\ref{eq:twentyseven}) required for a vortex unbinding transition.
This potentially divergent perturbation theory can be converted into
renormalization group recursion relations for effective couplings
$K(\ell)$ and $y(\ell)$ describing renormalized vortices with
effective core diameter $ae^{-\ell}$. These
differential equations read
\begin{eqnarray}
{dK^{-1}(\ell)\over d\ell}&=&
4\pi^2y^2(\ell)+O[y^4(\ell)]\;,
\eqnum{2.48{\rm a}}
\label{eq:fortyeighta}
\\
{dy(\ell)\over d\ell}&=&
[2-\pi K(\ell)]y(\ell)+
O[y^3(\ell)]\;.
\eqnum{2.48{\rm b}}
\label{eq:fortyeightb}
\end{eqnarray}
Another important result is the invariance of the
superfluid density along a renormalization group
trajectory,
\begin{equation}
K_R(K,y)=K_R(K(\ell),y(\ell))\;,
\eqnum{2.49}
\label{eq:fortynine}
\end{equation}
where $K_R(K,y)={\hbar^2\over m^2T}\rho_s^R(K,y)$.  This result
also follows from the Josephson scaling
relation for $\rho_s^R$ \cite{drnkos77}.

\myfigure{\epsfysize3in\epsfbox{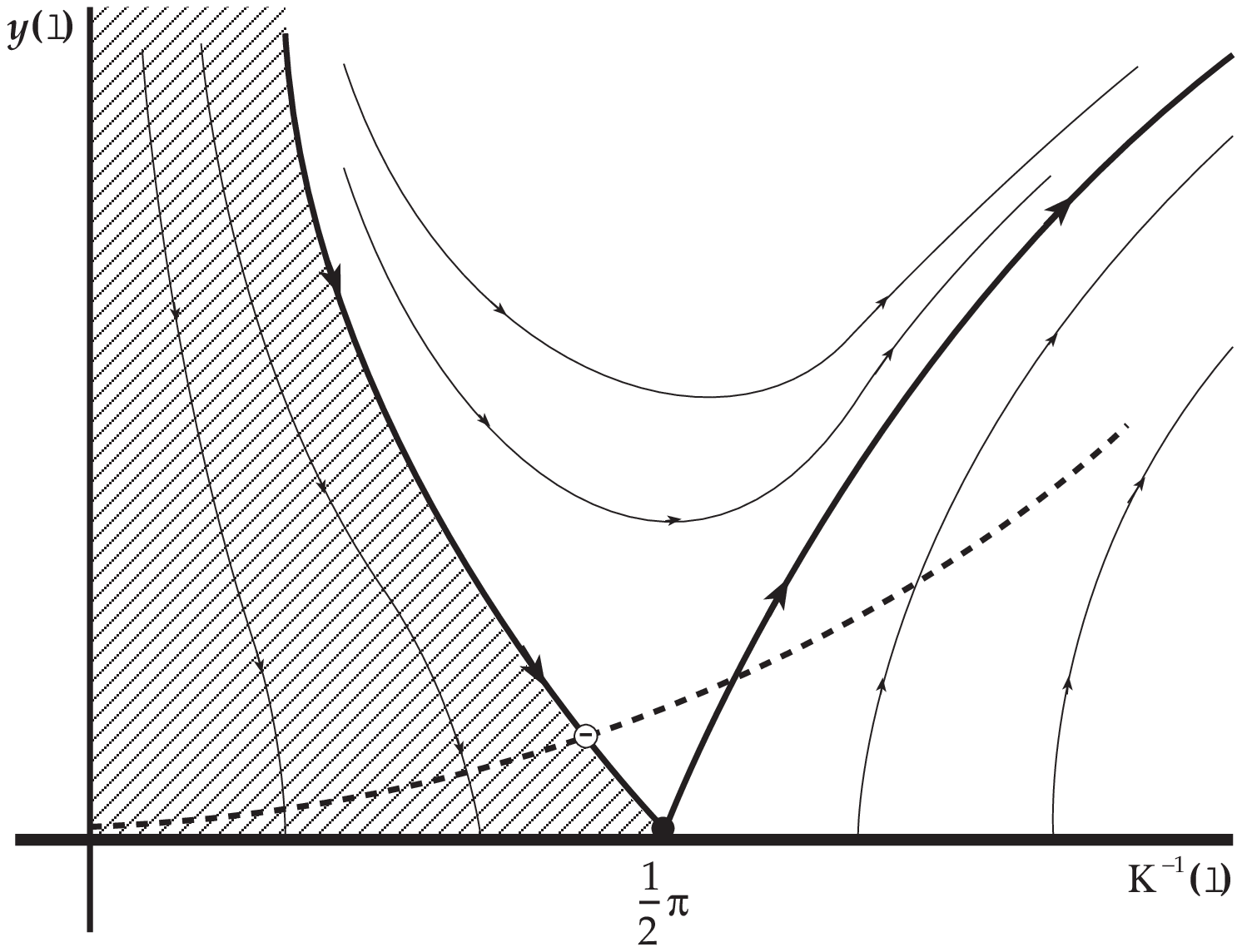}}{FIG.~4. \
Renormalization flows arising from the Kosterlitz recursion relations.
The shaded domain of attraction of the fixed line at $y(\ell)=0$ is a
superfluid.
A locus of initial conditions is shown as a dashed line. The superfluid
phase is bounded by the incoming separatrix which terminates at
$K^{-1}=2/\pi$.}

Equations (2.48) are the
famous Kosterlitz recursion relations, originally derived
by him using another method \cite{kos74}. The Hamiltonian
trajectories they generate in the $(K^{-1},y)$-plane are
shown in Fig.~4, together with a temperature-dependent
locus of initial conditions,
\begin{equation}
y_0=e^{-cK_0}\;,
\eqnum{2.50}
\label{eq:fifty}
\end{equation}
where $c$ is a constant.

Initial conditions to the left of the incoming separatrix
renormalize into the line of fixed points at $y=0$, which
describes the low-temperature phase. At higher temperatures,
$y(\ell)$ eventually becomes large, indicating that
vortices unbind at long wavelengths. One then expects
that order parameter correlations decay exponentially, as in
Eq. (\ref{eq:twentyone}), where the correlation length $\xi_+$
is related to the density of free vortices $n_f$:
\begin{equation}
n_f(T)\approx \xi_+^{-2}(T)\;.
\eqnum{2.51}
\label{eq:fiftyone}
\end{equation}

A variety of detailed predictions for the Kosterlitz-Thouless
transition follow from these recursion relations and the
transformation properties of various correlation functions under
the renormalization group \cite{kos74}. We focus here on the
prediction of a {\it universal} jump discontinuity in the
superfluid density \cite{drnkos77}. This is a direct
consequence of the relation (\ref{eq:fortynine}) and the
renormalization group flows shown in Fig.~4. Below
$T_c$, $y(\ell)$ tends to zero for large $\ell$, and we have
\begin{eqnarray}
K_R(K,y)&=&\lim_{\ell\rightarrow\infty}
K_R(K(\ell),y(\ell))\nonumber \\
&=&\lim_{\ell\rightarrow\infty} K(\ell)\quad
(T\;\alt\; T_c)\;.
\eqnum{2.52}
\label{eq:fiftytwo}
\end{eqnarray}
Since it is clear from Fig.~4 that this limit is just
$2/\pi$, at $T_c$, we have
\begin{equation}
\lim_{T\rightarrow T_c^-}
K_R(K,y)=\lim
_{T\rightarrow T_c^-}\;
{\hbar^2\rho_s^R(T)\over m^2T}=
{2\over\pi}\;,
\eqnum{2.53}
\label{eq:fiftythree}
\end{equation}
{\it independent} of the way in which the initial locus crosses the
incoming separatrix. In helium films of varying thickness on
varying substrates, one predicts a sequence of curves with jump
discontinuities and $T_c$ values all falling on a line with slope
\begin{equation}
{\rho_s(T_c^-)\over T_c^-}=
{2m^2\over\pi\hbar^2k_BT}
\approx 3.491\times 10^{-9}\;
g\;cm^{-2} K^{-1}\;.
\eqnum{2.54}
\label{eq:fiftyfour}
\end{equation}

\myfigure{\epsfysize3.5in\epsfbox{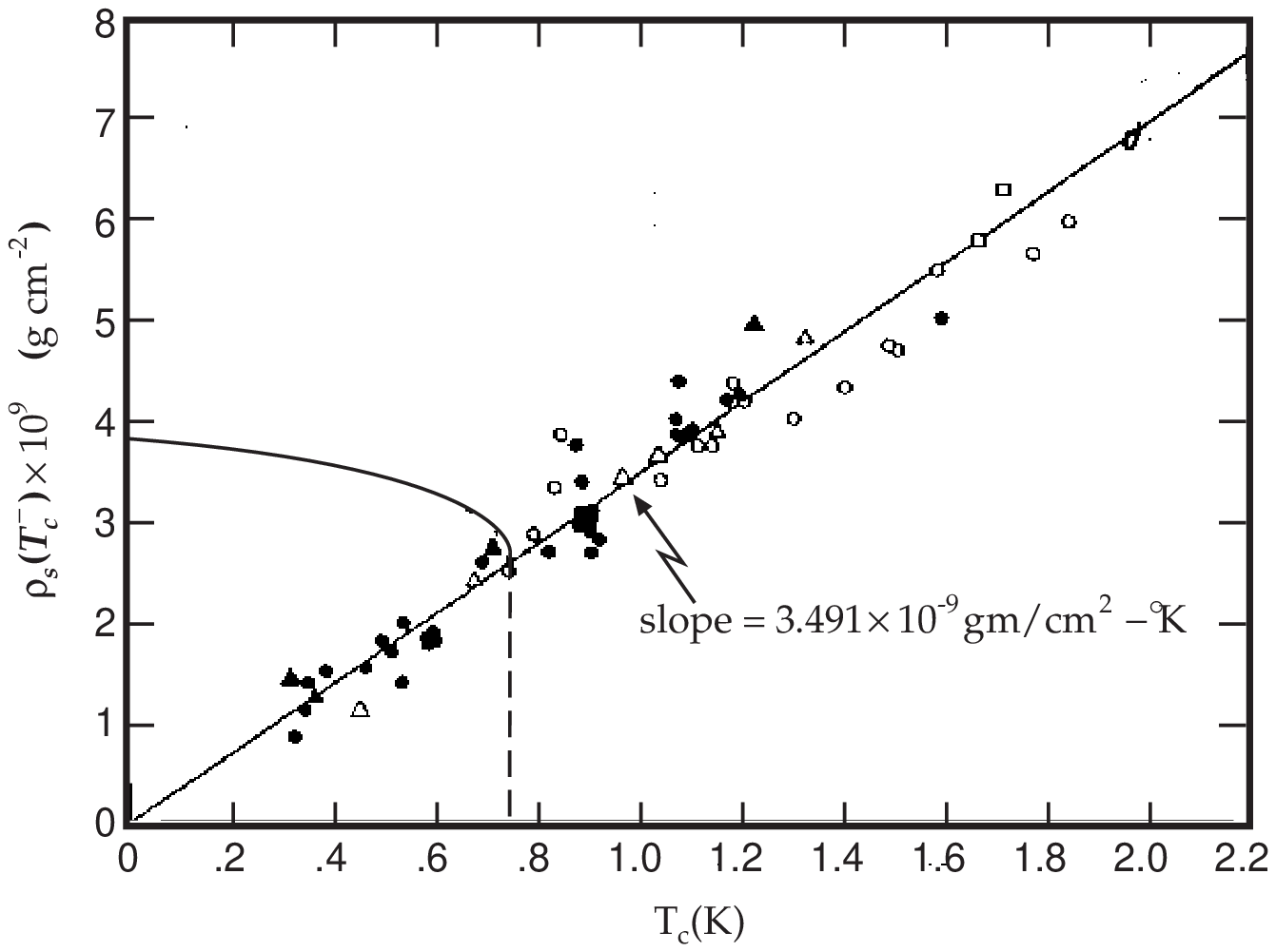}}{FIG.~5. \
Jump discontinuities in the superfluid density vs.\ temperature for
over 70 different experiments for  helium films. One representative
$\rho_s(T)$ curve is shown. Adapted  from Ref. \cite{bis78}.}

Fig.~5 shows the locus of jump discontinuities in the superfluid
density vs. temperature for over 70 experiments on helium films.
All points lie on a line with slope close to the universal
value (\ref{eq:fiftyfour}). This remarkable universal jump is
related to the universality of the critical exponent $\eta(T)$
at the Kosterlitz-Thouless transition. When vortices are taken
into account, $G(r)\sim {1\over r^{\eta(T)}}$ for $T<T_c$, provided
Eq. (\ref{eq:twenty}) is replaced by \cite{kos74}
\begin{equation}
\eta(T)=1/2\pi K_R(T)\;,
\eqnum{2.55}
\label{eq:fiftyfive}
\end{equation}
so that (neglecting logarithmic corrections) $\eta(T_c^-)=1/4$.

\subsection{Two-Dimensional Superconductors}
\label{subsec:twodim}

Superconducting films are similar to substrates coated with
superfluid He$^4$, with the understanding that the complex
order parameter $\psi(\vr)$ describes Cooper pairs of electrons.
Because the bosonic degrees of freedom are now charged, the
currents surrounding vortices are screened by a coupling to the
vector potential. The interactions between vortices are no longer
logarithmic at all distances, but instead die off rapidly at scales larger
than the London penetration depth. We first review a ``naive'' theory of
screening in two-dimensional superconductors. This theory is
``two-dimensional'' in the sense that all quantities are independent
of the $z$ coordinate. It would be directly applicable to
situations in which vortices are, in effect, infinitely
long rigid  rods in a {\it bulk} material, and no thermal
variation in the phase or vortex positions allowed along the
$z$ axis. The conclusion of the naive theory is that,
due to screening, vortices will always unbind at any nonzero
temperature \cite{kosthou72}. We then discuss real
superconducting films, which behave differently, due to the
spreading of the vortex magnetic field lines (see Fig.~6) as they emerge from
the top and bottom of the film \cite{pearl}. As emphasized by
Beasley {\it et al.} \cite{beasetal} vortices in sufficiently
thin superconducting films now interact logarithmically out to
an {\it effective} London penetration depth of order millimeters,
and the Kosterlitz-Thouless theory for neutral superfluids becomes
directly applicable over a wide range of length scales
in most samples. The detailed predictions of the
Kosterlitz-Thouless theory when transcribed to superconducting
films are discussed in Ref. \cite{bihdrn79}.

\myfigure{\epsfysize3in\epsfbox{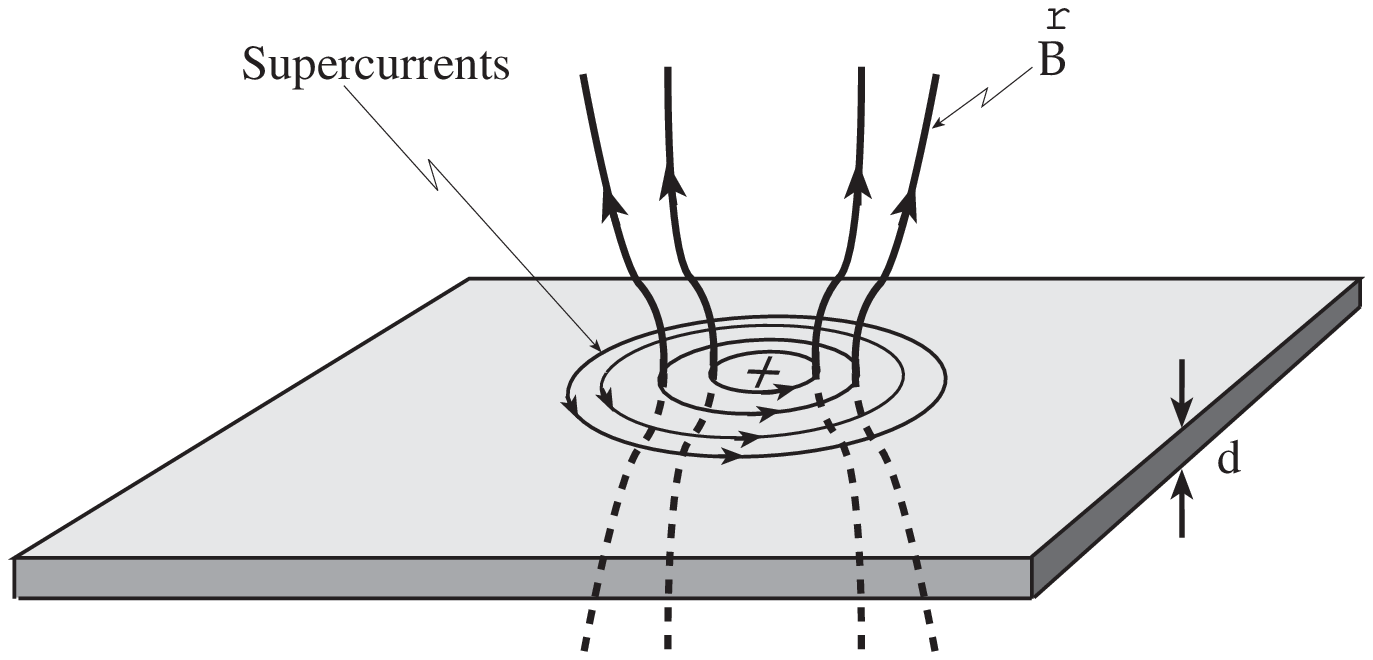}}{FIG.~6. \
Schematic of currents and magnetic field lines for a vortex in a
superconducting film. Currents only flow in a two-dimensional plane, so
the screening is weak.}

We begin with the ``naive'' theory of two-dimensional superconductors
because it provides a simple illustration of the
screening of defect interactions by a vector potential. A closely
related, but nonlinear screening phenomenon arises for dislocations
in membranes (see Sec.~III), and leads to the striking conclusion that
crystalline membranes must {\it always} melt into hexatic liquids at
any finite temperature.

\subsubsection{Naive Theory}
\label{subsubsec:naive}

We start with Ginzburg-Landau theory of superconductivity \cite{mtbook},
which generalizes Eq. (\ref{eq:eleven}) for neutral superfluids to
charged Cooper pairs,
\begin{eqnarray}
F&=&\int d^2r\left[
{1\over 2m^*}\left|
\left({\hbar\over i}\vnabla-{e^*\over c}\vec A\right)\psi
\right|^2 +{1\over 2} r|\psi|^2+u|\psi|^2\right. \nonumber \\
&\quad +&\left. {B^2\over 8\pi}-{1\over 4\pi} \vec H\cdot\vec B\right]
\;.
\eqnum{2.56}
\label{eq:fiftysix}
\end{eqnarray}
Here, $\psi(\vr)$ is a coarse-grained complex field which
represents the microscopic Cooper pair destruction operator ($\psi^*(\vec r)$
is
the coarse-grained pair creation operator),
$\vec A(\vr)$ is the vector potential, $e^*=2e$ is the pair
charge and $m^*=2m$ is the effective mass of the pair.
$\vec H$ is a constant external field, and $\vec B(\vr)=\vnabla
\times \vec A(\vr)$ is fluctuating local magnetic field.
Equation (\ref{eq:fiftysix}) is incorrect for real superconducting
films because it implicitly assumes that {\it both}
$\psi(\vr)$ and $\vec A(\vr)$ vanish outside the thin slab
occupied by the superconductor. This is a reasonable assumption
for $\vec \psi(\vr)$, but incorrect for $\vec B(\vr)$ and $\vec A(\vr)$.
This theory does, however, describe a bulk superconductor which is
infinitely stiff in the $\hat z$-direction, so that all quantities are
independent
of $z$. Equation (\ref{eq:fiftysix}) should then be interpreted as a
free energy per unit length. A related theory provides the correct description
of local
order in two-dimensional smectic liquid crystals \cite{tonerdrn}.

Let us see what Eq. (\ref{eq:fiftysix}) predicts for superconductors,
with the external magnetic field $\vec H$ set to zero. The modifications
introduced by the fluctuating magnetic field $\vec B(\vec r)$ outside the
sample will be discussed in the next subsection. Following our
treatment of helium films we assume that thermal excitations
have suppressed any possible ordering well below the mean field
transition temperature, so that $r\ll 0$. We again make the ``phase only''
approximation, and write
\begin{equation}
\psi(\vr)=\psi_0
e^{i\theta(\vr)}\;,
\eqnum{2.57}
\label{eq:fiftyseven}
\end{equation}
with $\psi_0=\sqrt{-r/4u}$. The two terms quadratic in $\psi(\vr)$ determine
the superconducting coherence length,
\begin{equation}
\xi_0=\sqrt{\hbar^2/m^*|r|}\;,
\eqnum{2.58}
\label{eq:fiftyeight}
\end{equation}
which is the length over which the order parameter rises from zero near
a vortex. On scales larger than $\xi_0$, so that the phase only
approximation is valid, free energy becomes
\begin{equation}
F={\rm const.} +
\int d^2r\left[{1\over 2}
\rho_s^0\left({\hbar\over m^*}\right)^2
\left|\vnabla\theta-{e^*\over\hbar c}\vec A
\right|^2+{1\over 8\pi}|\vnabla
\times \vec A|^2\right]\;,
\eqnum{2.59}
\label{eq:fiftynine}
\end{equation}
where
\begin{equation}
\rho_s^0=m^*|\psi_0|^2
\eqnum{2.60}
\label{eq:sixty}
\end{equation}
is the mass density of the Cooper pairs.

To see qualitatively how the vector potential affects the
behavior of vortices, imagine that the vortex solution (\ref{eq:twentythree})
for the phase variable is inserted in Eq. (\ref{eq:fiftynine}). Since
$|\vnabla\theta|\sim1/r$, this would lead to a logarithmically
diverging energy, if the vector potential were neglected. The vector
potential, however,  will cancel out this divergence and reduce the energy
provided it too falls off like $1/r$ with the correct coefficient
far from the vortex. This screening costs gradient energy, however, due
to the $|\vec\nabla\times\vec A|^2$ term. The two terms
of Eq.~(\ref{eq:fiftynine}) quadratic in $\vec A$ become
comparable when $\vec A(\vr)$ varies over a length scale $\lambda_0$,
with
\begin{equation}
\lambda_0=\sqrt{{(m^*c)^2\over
\rho_s^0 4\pi(e^*)^2}}\;.
\eqnum{2.61}
\label{eq:sixtyone}
\end{equation}
This scale is the London penetration depth, of order 1--10
thousand Angstroms in most superconductors. Because of this is the
scale over which screening sets in, we expect the vortex energy to be
finite and approximately equal to Eq. (\ref{eq:twentyfive})
with $R$ replaced by $\lambda_0$,
\begin{equation}
E_v\approx \pi\rho_s^0
\left({\hbar\over m^*}\right)^2
\ln(c\lambda_0/\xi_0)
\eqnum{2.62}
\label{eq:sixtytwo}
\end{equation}
where $c$ is an undetermined constant. Since the entropy of an
isolated vortex is still proportional to $\ln R$, and the energy is now
finite, we conclude that vortices will always be unbound in this
model \cite{kosthou72}. A more detailed analysis, similar to that
for finite energy defects like vacancies and interstitials in a
solid \cite{ashmer}, predicts a nonzero density of free vortices
\begin{equation}
\eta_f\approx\xi_0^{-2} e^{-E_v/T}
\eqnum{2.63}
\label{eq:sixtythree}
\end{equation}
at any temperature $T>0$.

Because Eq. (\ref{eq:fiftynine}) is a simple quadratic form, we can go beyond
the arguments sketched above, and determine the interactions between
screened vortices in more detail. As in helium films, functional integrals
weighted by $\exp[-F/T]$ will be dominated by vortex extrema of the
free energy. In  the Coulomb gauge, $\vnabla\cdot\vec A=0$, the
variational equation $\delta F/\delta\theta=0$ again leads to
$\nabla^2\theta=0$ away from the vortex cores, and we can immediately
write down the vortex contribution to $\vnabla\theta$ from $N$ vortices
at position $\{\vr_j\}$ with ``charges'' $\{s_j\}$,
\begin{equation}
\vnabla\theta(\vr)=
2\pi\sum_\alpha s_\alpha(\hz\times\vnabla)
G(\vr,\vr_\alpha)\;,
\eqnum{2.64}
\label{eq:sixtyfour}
\end{equation}
where $G(\vr,\vr_j)$ is the Green's function discussed in Section~II.C above.
Note from Eqs. (\ref{eq:thirtyone}) and (\ref{eq:thirtytwo}) that the curl
of $\vnabla\theta$ does not vanish,
\begin{equation}
\hz\cdot(\vnabla\times\vnabla\theta)=
2\pi\sum_{\alpha} s_\alpha\delta^{(2)}(\vr-\vr_\alpha)\;.
\eqnum{2.65}
\label{eq:sixtyfive}
\end{equation}

The vector potential is constrained by vortex singularities, even
though $\theta(\vec r)$ and $\vec A(\vec r)$ appear to decouple in
the Coulomb gauge. To see this, consider a closed path $\cal C$ bounding
an area $\Omega$ in the superconductor far compared to $\lambda_0$ from
any vortex cores. On such a path the first term in (\ref{eq:fiftynine})
has been minimized, insuring that $\vnabla\theta=
{e^*\over \hbar c}\vec A$ and that
\begin{equation}
\oint_{\cal C}\left(\vnabla\theta-{e^*\over \hbar c}\vec A
\right)\cdot d\vec \ell =0\;.
\eqnum{2.66}
\label{eq:sixtysix}
\end{equation}
It then follows from Eq. (\ref{eq:thirty}) that the magnetic flux
through $\Omega$ is given by the sum of the enclosed vortex ``charges,''
\begin{equation}
\mathop{\int\!\!\!\int}\limits_{\Omega} d^2r B_z=
\phi_0\sum_{\{r_\alpha\in\Omega\}}
s_\alpha\;,
\eqnum{2.67}
\label{eq:sixtyseven}
\end{equation}
where $\phi_0={2\pi \hbar c\over e^*}$ is the flux quantum.

A second variational equation follows from $\delta F/\delta \vec A=0$, i.e.,
\begin{equation}
\vnabla\times[\vnabla\times \vec A(\vr)]=
\vnabla\times\vec B(\vr)=
4\pi\rho_s^0\left(
{\hbar\over m^*}\right)^2
\left({e^*\over\hbar c}\right)
\left(\vnabla\theta-{e^*\over \hbar c}\right)
\vec A(\vr)\;.
\eqnum{2.68}
\label{eq:sixtyeight}
\end{equation}
Upon casting this equation in the form of Ampere's law
$\vnabla\times\vec B(\vr)=4\pi\vec J(\vr)/c$, we see that the
(gauge invariant) supercurrent associated with $\vec \theta$ and
$\vec A$ is
\begin{equation}
\vec J(\vr)=e^*|\psi_0|^2
{\hbar\over m^*}
\left(\vnabla\theta(\vr)-
{e^*\over\hbar c}\vec A(\vr)\right)\;.
\eqnum{2.69}
\label{eq:sixtynine}
\end{equation}
Note the close analogy with the formula for the superfluid {\it momentum}
density in helium films \cite{vinen},
\begin{equation}
\vec g(\vr)=m|\psi_0|^2
{\hbar\over m}\vnabla\theta(\vr)\;.
\eqnum{2.70}
\label{eq:seventy}
\end{equation}
A closed form equation for the magnetic field $B_z$ generated by a distribution
of vortices follows from taking the curl of Eq. (\ref{eq:sixtyeight}), and
using Eq. (\ref{eq:sixtyfive})
\begin{equation}
B_z-\lambda_0^2\nabla_\perp^2B_z=
\phi_0\sum_{\alpha=1}^N
s_\alpha\delta^{(2)}
(\vr-\vr_\alpha)\;,
\eqnum{2.71}
\label{eq:seventyone}
\end{equation}
where ``$\perp$'' denotes coordinates in the plane of the film.

Consider one isolated vortex with charge $s_\alpha$ at the origin.
The Fourier transformed field $B_z(\vq_\perp)$ which solves Eq.
(\ref{eq:seventyone})
is then
\begin{equation}
B_z(\vq)=
{s\phi_0\over 1+q_\perp^2\lambda_0^2}\;,
\eqnum{2.72}
\label{eq:seventytwo}
\end{equation}
which leads in real space to
\begin{equation}
B_z(r)={s\phi_0\over 2\pi\lambda_0^2}
K_0(r/\lambda_0)\;,
\eqnum{2.73}
\label{eq:seventythree}
\end{equation}
where $K_0(x)$ is the Bessel function, $K_0(x)\approx\ln(1/x)$, $x\ll 1$, and
$K_0(x)=\sqrt{\pi/2x}\;e^{-x}$, $x\gg 1$. The associated supercurrent
$\vec J(r)=\vnabla\times\vec B(r)$ is
\begin{equation}
\vec J(\vr)=
{sc\phi_0\over 8\pi^2\lambda_0^3}K_1(r/\lambda_0)\hat \theta\;,
\eqnum{2.74}
\label{eq:seventyfour}
\end{equation}
where $\hat\theta$ is a unit vector in the  azimuthal direction. Note that
$|\vec J(\vr)|\sim 1/r$ for $r\ll \lambda_0$, similar to the momentum
density in helium films. However, screening sets in for $r\gg \lambda_0$, and
$|\vec J(\vr)|\sim\exp(-r/\lambda_0)$. Similar manipulations lead
straightforwardly to the generalization of the vortex-free energy
(\ref{eq:fortyone}) for two-dimensional charged superfluids,
\begin{equation}
F_v=\epsilon_0
\sum_{i\not=j}
s_is_jK_0\left({|\vr_i-\vr_j|\over\lambda_0}\right)
+E_c\sum_j s_j^2\;,
\eqnum{2.75}
\label{eq:seventyfive}
\end{equation}
where $\epsilon_0=(\phi_0/4\pi\lambda_0)^2$ and $E_c$ is a vortex
core energy. Gaussian ``spin wave'' fluctuations about these vortex
extrema can also be included \cite{tonerdrn}. The finite range
of the interaction potential confirms our earlier conclusion that
vortices will unbind for entropic reasons at any finite temperature
in this model.

\subsubsection{Real Superconducting Films}

As illustrated in Fig.~6, the magnetic field lines generated by a
vortex in real superconducting films spread out into the empty
space above and below the plane of film. This spreading leads to
different behavior for the screening currents in the film than
that predicted by Eq.~(\ref{eq:seventyfour}). The true behavior
was first found by Pearl \cite{pearl}, and provides an interesting
illustration of how two-dimensional physics can be effected by the
outside, three-dimensional environment. We sketch the calculation here,
following the treatment of deGennes \cite{dege89}.

Start by integrating Eq. (\ref{eq:sixtynine}) over a film of thickness
$d$ along $\hat z$. Provided the film is much thinner than the London
penetration depth, we may assume that $\theta(\vr)$ and $\vec A_{2d}(\vr)$
remain constant across the film thickness. We denote the vector
potential by $\vec A_{2d}(\vr)$ to emphasize that this quantity is the
vector potential {\it in the two-dimensional film}, and not the full vector
potential $\vec A_{3d}(\vr,z)$ which describes the magnetic field in
all of three-dimensional space. Note that $\vr=(x,y)$ always
refers to coordinates in the two-dimensional plane perpendicular to
$\hat z$. If we imagine that all this current is concentrated in a
delta function sheet in the plane $z=0$, the full three-dimensional
current may be written
\begin{equation}
\vec J(\vr,z)=
{c\over 4\pi}{\phi_0\over 2\pi\lambda_{{\rm eff}}}
\left[\vnabla\theta(\vr)-
{2\pi\over\phi_0}\vec A_{2d}(\vr)\right]
\delta(z)\;,
\eqnum{2.76}
\label{eq:seventysix}
\end{equation}
where
\begin{equation}
\lambda\eff=\lambda^2/d\;.
\eqnum{2.77}
\label{eq:seventyseven}
\end{equation}
As we shall see, $\lambda\eff$ will play the role of an effective London
penetration depth in this problem.

The current (\ref{eq:seventysix}) provides a source for a {\it
three}-dimensional
magnetic field $\vec B(\vr,z)$ via Ampere's law, $\vnabla\times\vec B(\vr,z)=
4\pi\vec J(\vr,z)/c$. Upon setting $B=\vec \nabla\times\vec A_{3d}$ and using
the Coulomb gauge, Ampere's law becomes
\begin{equation}
-\nabla^2\vec A_{3d}(\vr,z)+
{1\over\lambda\eff}
\vec A_{2d}(\vr)\delta(z)=
{s\phi_0\over 2\pi\lambda\eff}
{\hat z\times\vr\over r^2}
\delta(z)\;,
\eqnum{2.78}
\label{eq:seventyeight}
\end{equation}
where $\nabla^2=\nabla_\perp^2+\partial^2/\partial z^2$ and we have inserted
the phase
gradient for a single vortex of charge $s$ at the origin. We now pass to
Fourier transformed vector potentials
\begin{equation}
\vec A_{3d}(\vq_\perp,q_z)=
\int d^2r\int dz e^{i\vq_\perp\cdot\vr}
e^{iq_zz}
\vec A_{3d}(\vr,z)
\eqnum{2.79}
\label{eq:seventynine}
\end{equation}
and
\begin{equation}
\vec A_{2d}(\vq_\perp)=
\int d^2r e^{i\vec q_\perp\cdot\vr}
\vec A_{2d}(\vr)
\eqnum{2.80}
\label{eq:eighty}
\end{equation}
and find that (\ref{eq:seventyeight}) becomes
\begin{equation}
\vec A_{3d}(\vq_\perp, q_z)
+{1\over\lambda\eff(q_z^2+q_\perp^2)}
\vec A_{2d}(\vq_\perp)=
{is\phi_0\over\lambda\eff(q_z^2+q_\perp^2)}
{\hat z\times\vq_\perp\over q_\perp^2}\;.
\eqnum{2.81}
\label{eq:eightyone}
\end{equation}

The $2d$ and $3d$ vector potentials must agree on the plane
$z=0$, i.e.,  $\vec A_{3d}(\vr,z=0)=\vec A_{2d}(\vr)$, so that
\begin{equation}
\vec A_{2d}(q_\perp)=
\int_{-\infty}^\infty
{dq_z\over 2\pi}\vec A_{3d}
(\vq_\perp,q_z)\;.
\eqnum{2.82}
\label{eq:eightytwo}
\end{equation}
After integrating Eq. (\ref{eq:eightyone}) over $q_z$ and using this
result we can solve for
$\vec A_{2d}(\vec q_\perp)$,
\begin{equation}
\vec A_{2d}(\vq_\perp)=
{is\phi_0(\hat z\times\vq_\perp)\over
q_\perp^2(1+2\lambda\eff q_\perp)}\;.
\eqnum{2.83}
\label{eq:eightythree}
\end{equation}
Upon inserting Eq. (\ref{eq:eightythree}) into the Fourier transform of Eq.
(\ref{eq:seventysix})
after setting $\vec J(\vr,z)\equiv d\vec  J_{2d}(\vr)\delta(z)$, we have
\begin{equation}
\vec J_{2d}(\vq_\perp)=
{c\over 2\pi}\;
{is\phi_0(\hat z\times\vq_\perp)\over
dq_\perp(1+2\lambda\eff q_\perp)}
\eqnum{2.84}
\label{eq:eightyfour}
\end{equation}
and so the superconducting current in real space is
\begin{equation}
\vec J_{2d}(\vr)=
{is\phi_0\over 2\pi d}
\int{d^2q_\perp\over (2\pi)^2}
{e^{-i\vq_\perp\cdot \vec r}\over
(1+2\lambda\eff q_\perp)}
{\hat z\times\vq_\perp\over q_\perp}\;.
\eqnum{2.85}
\label{eq:eightyfive}
\end{equation}

The behavior of this screening current depends on the effective London
penetration depth. For $r\ll \lambda\eff$, we find
\begin{equation}
\vec J_{2d}(\vr)=
{sc\phi_0\over 8\pi^2 d\lambda\eff r}
\hat \theta
\eqnum{2.86}
\label{eq:eightysix}
\end{equation}
which agrees with the $r\ll \lambda_0$ limit of Eq. (\ref{eq:seventyfour}) for
our
``naive'' model. This unscreened $1/r$ falloff now continues, however, out to a
much larger distance $\lambda\eff=(\lambda_0/d)\lambda_0\gg \lambda_0$. For
$r\gg \lambda\eff$, the limiting behavior of (\ref{eq:eightyfive}) is
\begin{equation}
\vec J_{2d}(\vr)=
{sc\phi_0\over 4\pi^2dr^2}
\hat\theta\;.
\eqnum{2.87}
\label{eq:eightyseven}
\end{equation}
The exponential screening of vortex currents in the ``naive''
model is thus replaced by a $1/r^2$ power-law falloff.

Similar results hold for the interactions between vortex
pairs \cite{dege89}. Vortices interact logarithmically, as into
neutral superfluids for $r\ll\lambda\eff$, but exhibit a weaker
$1/r$ potential for $r\gg\lambda\eff$. Since $\lambda\eff$ can be
of order fractions of a centimeter for film thicknesses $d=10-100\;\AA$, the
Kosterlitz-Thouless theory becomes directly applicable on essentially all
length scales for sufficiently thin films \cite{beasetal}. The
relatively weak screening compared to bulk systems arises because
currents are confined to a thin plane instead of  forming rings along
the entire $z$ axis.

\section{Defects in Membranes and Monolayers}

Two-dimensional crystals have much in common with two-dimensional
superfluids. There are many important experimental examples,
including rare gases adsorbed onto periodic substrates
like graphite, Langmuir-Blodgett films of ampiphillic molecules
at air-water interfaces, freely suspended liquid crystal films,
electron layers trapped at the surface of liquid helium, and
assemblies of colloidal particles confined between two glass
plates \cite{drn83,bondor}. Crystals consisting of a few atomic or molecular
layers display algebraic decay of a translational order parameter,
similar to Eq. (\ref{eq:nineteen}), and the crystal elastic
constants play a role similar to that of the superfluid
density \cite{drn83}. As emphasized by Kosterlitz and
Thouless \cite{kosthou72} and by Berezinski \cite{bere70},
dislocations in such crystals are point defects with a logarithmically
diverging energy as a function of system size, and one might expect
them to melt via a dislocation unbinding mechanism.

When the detailed dislocation unbinding theory was worked out
\cite{bihdrn78,young},
there was a surprise. Melting via dislocations leads not to an isotropic
liquid, as proposed originally \cite{kosthou72,bere70}, but produces
instead a new hexatic phase of matter, with residual bond orientational
order \cite{bihdrn78}. The long-range bond orientational order in
two-dimensional
crystals at low temperatures is converted into algebraically decaying
correlations in a hexatic fluid by a gas of unbound dislocations.
Each dislocation, moreover, contains an embryonic pair of
orientational defects called disclinations in its core. These disclinations
separate and interact logarithmically in the hexatic phase, and
themselves unbind via a second phase transition at sufficiently high
temperatures. The latent heat associated with the usual first-order
melting point can thus be spread out over an entire intermediate phase,
separated from the low-temperature crystal and high-temperature liquid
by two continuous phase transitions.

We call the experimental systems mentioned above ``monolayers,'' to emphasize
that they are constrained to be approximately flat. In all cases, there are
nearby substrates, walls or interfaces which force the degrees of
freedom to ``layer,'' i.e., to lie a plane. Excitations out of the
plane are strongly disfavored by, for example, a surface tension. There is,
however,
another important class of two-dimensional materials called ``membranes''
\cite{drnwein}. Membranes are two-dimensional associations of molecules
different from the three-dimensional fluid medium in which they are
embedded. Examples include lipid bilayers in water \cite{drnwein} or
spectrin protein skeletons extracted from red blood cells \cite{schmi}.
Because membranes are not confined to an interface between two different
phases, the surface tension vanishes and they exhibit wild fluctuations
out the plane while retaining a local two-dimensional topology.

In this section, we review the physics of defects such as
dislocations, disclinations, grain boundaries, vacancies and
interstitials in membranes and monolayers. We discuss crystalline
monolayers first, and then show how defect buckling screens
defect energies in membranes \cite{drnpel,seungdrn,cardrn}. The buckling
transition of disclinations in hexatic membranes \cite{seung} is also
described. We point out a fundamental  asymmetry in the populations of
positive and negative disclinations in liquid
membranes with free boundary conditions or with fluctuating
topologies. For a discussion of the spin-glass-like
statistical mechanics of {\it polymerized} membranes with
{\it quenched} random distributions of defects such as impurity
atoms, grain boundaries, see Refs.~\cite{radzdrn,morselub1,drnrad}.

\subsection{Landau Theory and Elasticity of Tethered Membranes}

Consider first the statistical mechanics of two-dimensional assemblies of
atoms and molecules {\it without} defects. To exclude
defects, we assume a perfect triangular lattice of monomers embedded in
three dimensions and {\it tethered} together via an unbreakable
network of covalent bonds. Such ``tethered surfaces'' \cite{kantetal,kantdrn}
exhibit interesting fluctuations and phase transitions even in the absence
of defects. Once in the low temperature, broken symmetry phase of the relevant
Landau theory, defects can be introduced for both monolayers and membranes.

Consider first the high temperature, crumpled phase of a tethered surface,
similar to the crumpled state of a linear polymer chain. We describe the
membrane by a function $\vr(x_1,x_2)$, where $\vr$ is a
three-dimensional vector specifying the position of the monomers as a
function of two internal coordinates $x_1$ and $x_2$ fixed to the
monomers. It turns out that the crumpled membrane can undergo a
spontaneously symmetry-breaking into a flat phase below a crumpling
temperature $T_c$ \cite{drnpel}. The order
parameters for this transition are the surface tangents $\{\vec
t_\alpha=d\vr/dx_\alpha
\;,\;\alpha=1,2\}$
and the Landau free energy describing the transition is \cite{pacz},
\FL
\begin{eqnarray}
F[\vr(x_1,x_2)]=&\int& d^2x\left[{1\over 2}\kappa(\partial_\alpha^2\vr)^2
\right.\nonumber \\
&+& \left.{1\over 2}a(\partial_\alpha \vr)^2+b(\partial_\alpha\vr
\cdot\partial_\beta\vr)^2+
c(\vec\partial\gamma\vr\cdot\partial\gamma\vr)^2
+\cdots\right]
\eqnum{3.1}
\label{eq:tone}
\end{eqnarray}
Within this expansion in the tangents, the probability of a surface
configuration $\vr(x_1,x_2)$ is proportional to $\exp\{-F[\vr(x_1,x_2)]/T\}$.
Note that $F$ is invariant under translations and rotations both within the
embedding space and in the internal coordinates, as it should be.
Self-avoidance between monomers at very different values of $\vec x=
(x_1,x_2)$, but close together in the embedding space is neglected,
although it is not hard to incorporate this into the model
\cite{pacz}. We assume that $a$ changes sign at the mean field
crumpling transition, $a=a'(T-T_c)$, just as in the more
conventional Landau theories for superfluids discussed in Sec.~II.

We discuss here only the low-temperature flat phase. When $a<0$, $F$ is
minimized by
\begin{eqnarray}
\vr_0&=&
\langle\vr(x_1,x_2)\rangle\nonumber \\
&=& m[x_1\vec e_1+x_2\vec e_2]\;,
\eqnum{3.2}
\label{eq:ttwo}
\end{eqnarray}
where $\vec e_1$ and $\vec e_2$ are an arbitrary orthogonal pair of unit
vectors in the three-dimensional embedding space and
\begin{equation}
m={1\over 2}\sqrt{{-a\over b+2c}}\;.
\eqnum{3.3}
\label{eq:tthree}
\end{equation}
Note that $\langle \vec t_\alpha\rangle=\langle d\vr/dx_\alpha\rangle=m\vec
e_\alpha$,
so that $m$ is the ``amplitude'' of the tangent order parameters. The average
of the surface tangents vanishes in the high-temperature phase, because the
surface is highly crumpled. Additional physical insight is provided if we
rewrite the free energy as
\FL
\begin{eqnarray}
F={\rm const.}+\int
d^2x\left\{{1\over 2}\kappa(\partial_\alpha^2\vr)^2\right. &+&
\mu\left[{\partial\vr\over\partial x_\alpha}\cdot
{\partial\vr\over\partial x_\beta}-m^2\delta_{\alpha\beta}\right]^2
\nonumber \\
&+&
\left.\lambda\left[{\partial\vr\over\partial x_\gamma}\cdot
{\partial\vr\over\partial x_\gamma}-2m^2\right]^2\right\}
\eqnum{3.4}
\label{eq:tfour}
\end{eqnarray}
where
\begin{equation}
\mu=4 bm^4
\eqnum{3.5{\rm a}}
\label{eq:tfivea}
\end{equation}
and
\begin{equation}
\lambda=8 cm^4\;.
\eqnum{3.5{\rm b}}
\label{eq:tfiveb}
\end{equation}
This free energy is a sum of a bending energy, controlled by $\kappa$,
and stretching contributions governed by elastic moduli $\mu$ and $\lambda$.
The stretching terms provide an energetic penalty whenever the induced
metric in the embedding space,
\begin{equation}
g_{\alpha\beta}(x_1,x_2)=
{\partial\vr\over\partial x_\alpha}\cdot
{\partial\vr\over\partial x_\beta}\;,
\eqnum{3.6}
\label{eq:tsix}
\end{equation}
deviates from a flat background value
\begin{equation}
g_{\alpha\beta}^0=m^2\delta_{\alpha\beta}\;.
\eqnum{3.7}
\label{eq:tseven}
\end{equation}

The low energy Goldstone modes associated with the flat phase are phonons.
To study these excitations, we proceed as in the ``phase only''
approximation for superfluids \cite{eq2-12} and set
\begin{equation}
\vr(x^1,x^2)\approx m(x_1+u_1)\vec e_1
+m(x_2+u_2)\vec e_2+f\vec e_3\;,
\eqnum{3.8}
\label{eq:teight}
\end{equation}
where $\vec u(x_1,x_2)$ is an in-plane phonon field and $f(x_1,x_2)$ is an
out-of-plane displacement along $\vec e_3=vec e_1\times \vec e_2$.
With the neglect of an  additive constant,
the free energy (\ref{eq:tthree}) becomes
\begin{eqnarray}
F&=&\int d^2x\left\{ {\mu\over 4}
[\partial_\alpha u_\beta+\partial_\beta u_\alpha+(\partial_\alpha f)
(\partial_\beta f)]^2\right.
\nonumber \\
&\quad +& {\lambda\over 2}[\partial_\gamma u_\gamma+{1\over 2}
(\partial_\gamma f)^2]^2
+{\kappa\over 2}\int d^2x(\nabla^2f)^2\;.
\eqnum{3.9}
\label{eq:tnine}
\end{eqnarray}
This expression is identical to the energy of a bent elastic plate,
where $\kappa$ is bending rigidity, $\mu$ is a shear modulus, and
$\mu+\lambda$ is the bulk modulus \cite{lanlif}.
One often finds the stretching energy contributions written in terms
of the nonlinear strain matrix $u_{\alpha\beta}(x_1,x_2)$
\begin{equation}
u_{\alpha\beta}=
{1\over 2}(\partial_\alpha u_\beta+\partial_\beta u_\alpha)+
{1\over 2}(\partial_\alpha f)(\partial_\beta f)\;.
\eqnum{3.10}
\label{eq:tten}
\end{equation}
We have neglected terms nonlinear in $\vec u$, which are less important than
the
nonlinear terms in $f$ which we have kept. It can be shown that the
nonlinear terms in $f$ cause the renormalized bending rigidity to {\it diverge}
at
long wavelengths, leading to the remarkable conclusion that the low-temperature
flat phase represents a genuine broken continuous symmetry with true long-range
order \cite{drnwein,drnpel}. Broken continuous symmetries two-dimensional
systems at finite temperatures
are usually impossible, and are replaced instead by exponential decay
of correlations or by algebraically decaying order such as that found for
superfluid helium films in Eq. (\ref{eq:nineteen}).

Note the similarity between Eq. (\ref{eq:tnine}) the free energy Eq.
(\ref{eq:fiftynine}) which arose in our ``naive'' theory of
superconducting films. The phase gradient $\vnabla\theta$ is replaced by the
linearized strain matrix ${1\over 2}(\partial_\alpha u_\beta+\partial_\beta
u_\alpha)$.
The superfluid density is replaced by the elastic constants $\mu$ and
$\lambda$. The role of the ``vector potential'' is played by gradients
of $f$, and the bending rigidity contribution replaces the field
energy term ${1\over 8\pi}|\vnabla\times\vec A|^2$ for superconductors.
Because ``vector potential'' contribution to Eq. (\ref{eq:tten}) is
nonlinear, the two theories are certainly not identical. Nevertheless,
we shall find striking similarities when we relax the constraint of
perfect sixfold coordination and allow for defects: If we
first set $f=0$, then Eq. (\ref{eq:tnine}) can be used to calculate defect
energies in planar ``monolayers''---see below. Defects in monolayers are
like the vortex singularities discussed earlier for neutral superfluids.
When $f\not= 0$, however, the theory applies to the low-temperature
phase of ``membranes.'' As we shall see, these defects are then
screened by nonzero gradients of $f$, similar to the screening of
vortices by the vector potential in superconductors.
The corresponding reduction in energy for membranes is accompanied by
defect buckling out of the plane defined by $f=0$.

\subsection{Defects in Monolayers}

Fig.~7 shows a typical particle configuration for a thermally excited
monolayer in its crystalline phase. Most particles, indicated by
circles, have six nearest neighbors, as determined by the Dirichlet
or Wigner-Seitz construction: The nearest neighbor coordination number of a
particle
is the number of edges of the minimal polygon formed by the bisectors of the
lines connecting it to its near neighbors. The 5- and 7-fold
coordinated particles (indicated by diamonds and asterisks, respectively)
are orientational disclination defects in the otherwise 6-fold
coordinated triangular lattice. An elementary application of Euler's
theorem shows that the average coordination number
with periodic boundary conditions must be exactly six \cite{runnels}, so
the numbers of 5's and 7's must be equal. A high-temperature liquid can
be viewed as a dense plasma of disclinations. The ``plus'' and
``minus'' disclinations (5's and 7's) in a liquid annihilate and
pair up with decreasing temperature to form a crystal.

\myfigure{\epsfysize3.5in\epsfbox{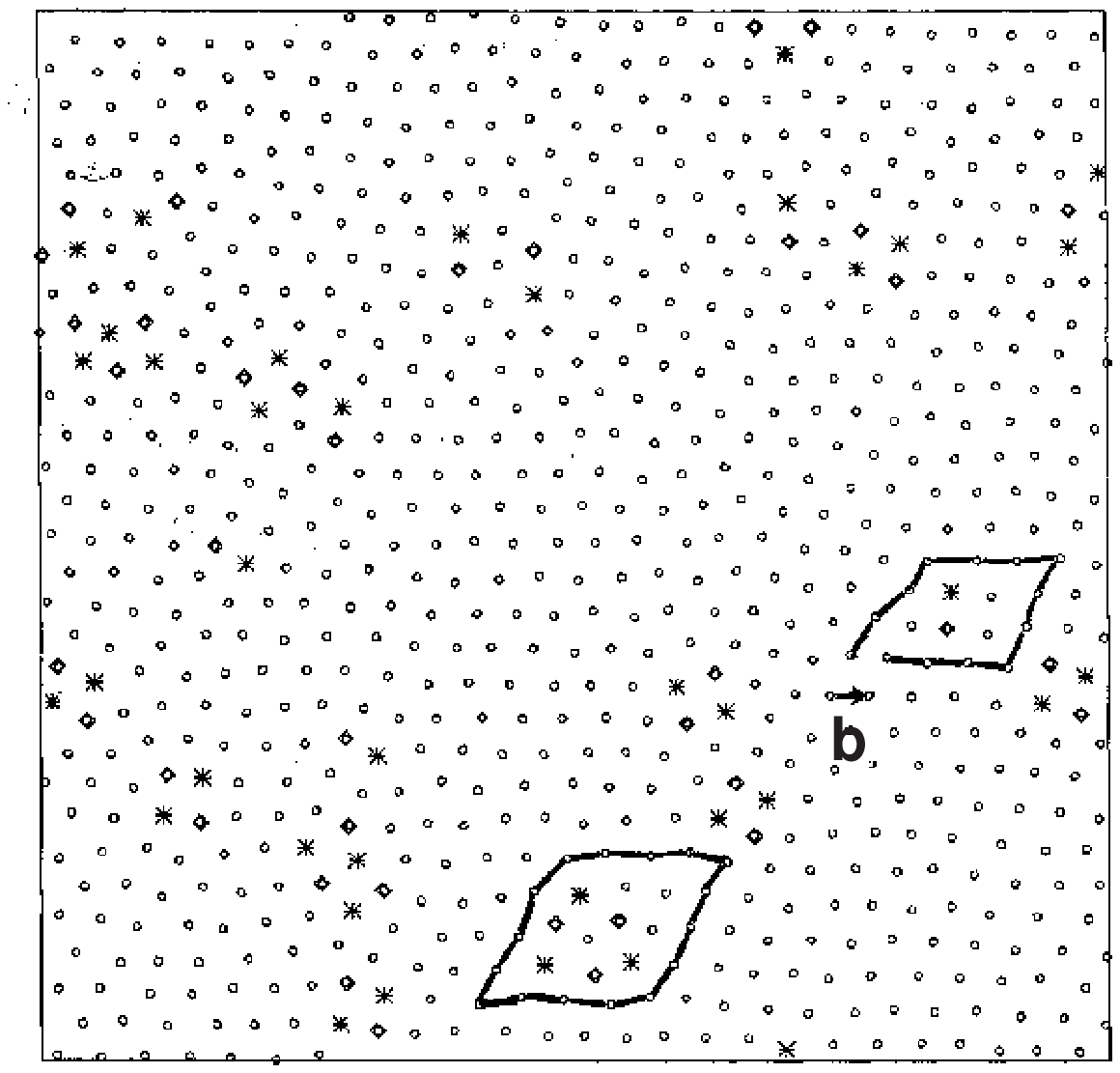}}{FIG.~7. \
Snapshot of a particle configurations in a computer simulation of a
``monolayer.'' Six-fold coordinated particles are shown as circles.
Five-fold and seven-fold disclination defects are shown as diamonds and
asterisks,
respectively. Heavy contours surround a dislocation and an
interstitial. From a simulation of classical electrons interacting with a
repulsive $1/r$ potential by Rudolph Morf.}

Disclination pairing is very evident in Fig.~7. An isolated
disclination pair is, in fact, a dislocation defect. As illustrated
in the figure, dislocations are characterized by the amount a path
which would close on a perfect lattice fails to close. This mismatch,
or ``Burgers vector,'' is a lattice vector  of the
underlying triangular crystal. It acts like a discrete
vector ``charge'' attached to the dislocation, is independent of the
exact contour chosen and points at right angles to a line connecting
the 5 to the 7 in the dislocation core. More generally, such circuits
determine the vector sum of all dislocation Burgers vectors contained
inside. The Burgers circuit does in fact close for the other
circuit shown in Fig.~7. The necklace of alternating
5's and 7's it contains can be viewed as three dislocations with radial Burgers
pointing at 120 degree angles to each other. It represents a third type
of defect, an interstitial. To see this, note that since the closed
Burgers parallelogram is 4 lattice units on a side, it would enclose
$3\times 3=9$ particles if the lattice were perfect. In fact, there are
10 particles inside the contour, indicating the presence of an extra
atom, or interstitial. The presence of a vacancy, the
``antidefect'' of the interstitial, could be detected by similar means.

To calculate the energies of the various monolayer defects discussed
above, we use the continuum elastic free energy (\ref{eq:tnine}) with
$f=0$,
\begin{equation}
F={1\over 2}\int d^2r[2\mu u_{ij}^2+\lambda u_{kk}^2]
\eqnum{3.11}
\label{eq:televen}
\end{equation}
where
\begin{equation}
u_{ij}={1\over 2}
(\partial_iu_j+\partial_ju_i)\;.
\eqnum{3.12}
\label{eq:ttwelve}
\end{equation}
In experimental monolayer systems, the constraint $f=0$ might be imposed
by a strong wall or substrate potential. For particles confined at an
interface with a surface tension $\sigma$, we should add a term
\begin{equation}
\delta F=
{1\over 2}\sigma\int d^2x
(\vnabla f)^2
\eqnum{3.13}
\label{eq:tthirteen}
\end{equation}
to Eq. (\ref{eq:tnine}). Upon integrating out the $f$ field in
perturbation theory, and we recover a free energy of the form
(\ref{eq:televen}),
with renormalized elastic constants $\mu$ and $\lambda$ \cite{sachdrn}.

Dislocations, disclinations and other defects can be introduced into
the theory in a way similar to the discussion of superfluid
vortices in Sec.~II. First, however, we determine the equations satisfied by
free energy extrema away from defect cores. The variation of the monolayer
free energy (\ref{eq:televen}) with respect to $\vec u$ leads to
\begin{equation}
\partial_i\sigma_{ij}=0\;,
\eqnum{3.14}
\label{eq:tfourteen}
\end{equation}
where
\begin{equation}
\sigma_{ij}=2\mu u_{ij}+
\lambda u_{kk}\delta_{ij}
\eqnum{3.15}
\label{eq:tfifteen}
\end{equation}
is the stress tensor. Equation (\ref{eq:tfourteen}) will be satisfied
automatically if we introduce the Airy stress function $\chi$ via
\begin{equation}
\sigma_{ij}(\vr)=
\epsilon_{im}\epsilon_{jn}
\partial_m\partial_n\chi(\vr)\;.
\eqnum{3.16}
\label{eq:tsixteen}
\end{equation}
The individual components of $\sigma_{ij}$ are thus
\begin{equation}
\sigma_{xx}={\partial^2\chi\over\partial y^2}
\;,\quad
\sigma_{yy}={\partial^2\chi\over\partial x^2}
\;,\quad
\sigma_{xy}=-{\partial^2\chi\over\partial x\partial y}
\;.
\eqnum{3.17}
\label{eq:tseventeen}
\end{equation}
The function $\chi(\vr)$ is similar to the vector potential one uses to
insure $\vnabla\cdot\vec B=0$ in Maxwell's equations.
If we are able to find $\chi$, we know $\sigma_{ij}$, and hence the
strain matrix by inverting Eq. (\ref{eq:tfifteen}),
\begin{eqnarray}
u_{ij}&=& {1\over 2\mu}\sigma_{ij}-
{\lambda\over 4\mu(\mu+\lambda)}\sigma_{kk}
\delta_{ij}\nonumber \\
&=& {1\over 2\mu}
\epsilon_{im}\epsilon_{jn}
\partial_m\partial_n\chi-
{\lambda\delta_{ij}\over
4\mu(\mu+\lambda)}
\nabla^2\chi\;.
\eqnum{3.18}
\label{eq:teighteen}
\end{eqnarray}

So far, we have not used the important fact that $u_{ij}$ is determined
by gradients of a displacement field via Eq. (\ref{eq:ttwelve}).
The corresponding requirement on the superfluid velocity, as
given by Eq. (\ref{eq:fourteen}), is
\begin{equation}
\vnabla\times\vv_s=0\;,
\eqnum{3.19}
\label{eq:tnineteen}
\end{equation}
which must hold away from vortex cores. The analogous compatibility
condition on $u_{ij}$ follows from applying the operator
$\epsilon_{ik}\epsilon_{j\ell}\partial_k\partial_\ell$ to both
sides of Eq. (\ref{eq:teighteen}). This operator vanishes when acting
on $u_{ij}(\vr)$, provided various derivatives of the displacement
field commute. Equation (\ref{eq:teighteen}) then simplifies to give a
biharmonic equation of $\chi(\vr)$,
\begin{eqnarray}
{1\over K_0}\nabla^4\chi(\vr)&=&{1\over 2}
\epsilon_{ij}\epsilon_{j\ell}\partial_k\partial_\ell
(\partial_iu_j+\partial_ju_i)\nonumber \\
&=& 0 \;,
\eqnum{3.20{\rm a}}
\label{eq:ttwentya}
\end{eqnarray}
with
\begin{equation}
K_0={4\mu(\mu+\lambda)\over
2\mu+\lambda}\;.
\eqnum{3.20{\rm b}}
\label{eq:ttwentyb}
\end{equation}

\myfigure{\epsfysize3.5in\epsfbox{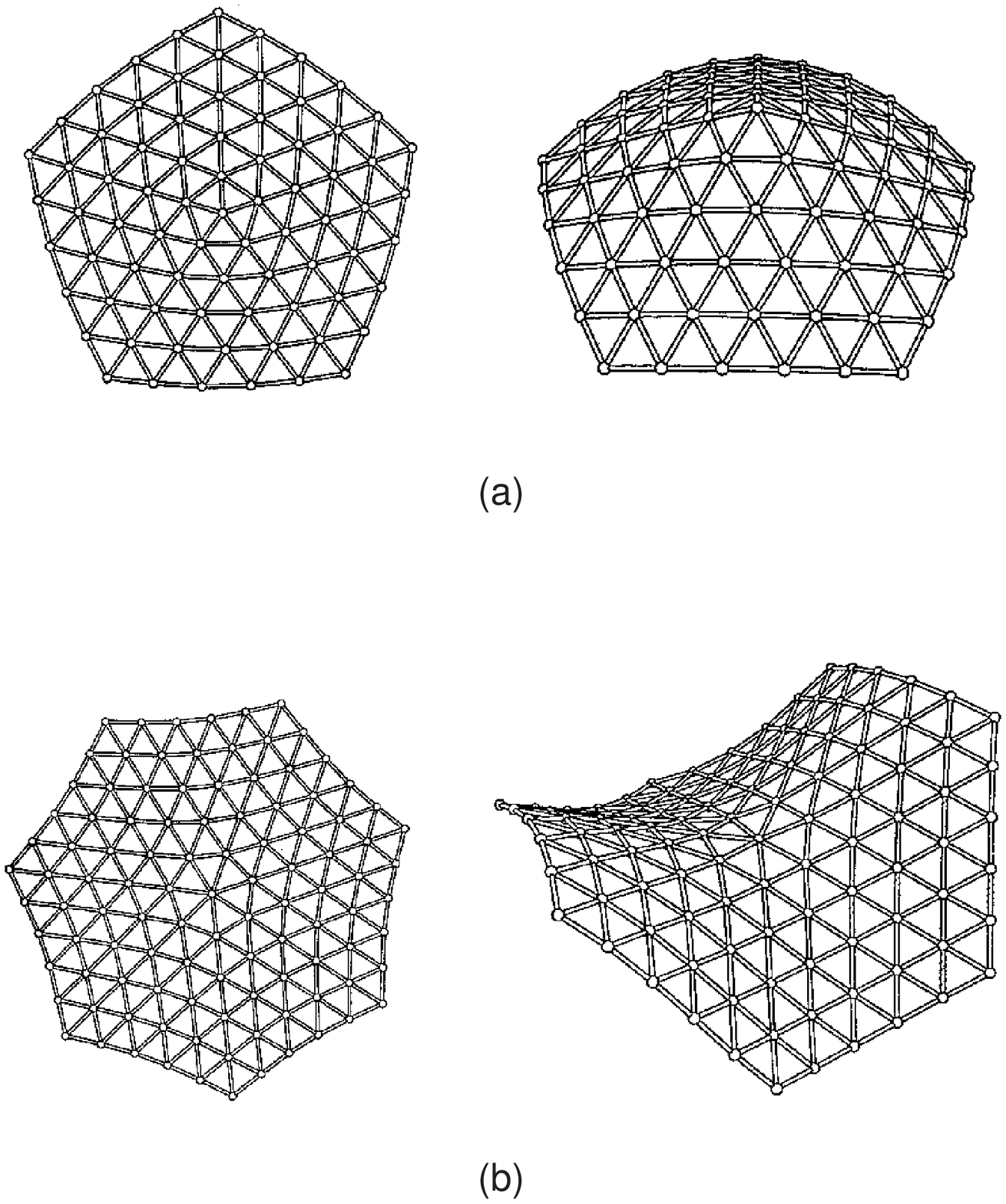}}{FIG.~8. \
Five- (a) and seven-fold (b) disclinations in   monolayers (flat)
and membranes (buckled). From Ref.~[37].}

Equation (3.20) must be satisfied ``almost everywhere,'' that is away
from defect cores. Defects introduce source terms on the right-hand
side, similar to the vortex density which appears in Eq. (\ref{eq:thirtyfour}).
As for superfluid vortices, defects represent points where
derivatives fail to commute. Because the line integral of the differential
bond angle field $\theta(\vr)$ around a disclination in a triangular lattice
must be an integral multiple of $2\pi/6=60^\circ$,
\begin{equation}
\oint_{\cal C} d\theta(\vr)=s
{2\pi\over 6}\;,
\eqnum{3.21}
\label{eq:ttwentyone}
\end{equation}
we have
\begin{equation}
\epsilon_{ij}\partial_i\partial_j\theta(\vr)=
{\pi\over 3}\sum_\alpha
s_\alpha\delta(\vr-\vr_\alpha)
\eqnum{3.22}
\label{eq:ttwentytwo}
\end{equation}
for a collection of disclinations with ``charges'' $\{s_\alpha=\pm 1, \pm 2,
\ldots\}$
at positions $\{\vr_\alpha\}$. Isolated 5- and 7-fold disclinations are
shown in Fig.~8. The local bond angle $\theta(\vr)$ in a crystalline solid is
given by the antisymmetric part of the strain tensor,
\begin{eqnarray}
\theta(\vr)&=&{1\over 2}
[\partial_xu_y(\vr)-\partial_yu_x(\vr)]\nonumber \\
&=& {1\over 2} \epsilon_{k\ell}\partial_k u_\ell\;,
\eqnum{3.23}
\label{eq:ttwentythree}
\end{eqnarray}
so the noncommunitivity of derivatives for disclinations takes the form,
\begin{equation}
{1\over 2} \epsilon_{ij}\epsilon_{k\ell}\partial_i\partial_j
\partial _ku_\ell=
{\pi\over 3}
\sum_\alpha
s_\alpha\delta(\vr-\vr_\alpha)\;.
\eqnum{3.24}
\label{eq:ttwentyfour}
\end{equation}

\myfigure{\epsfysize3.5in\epsfbox{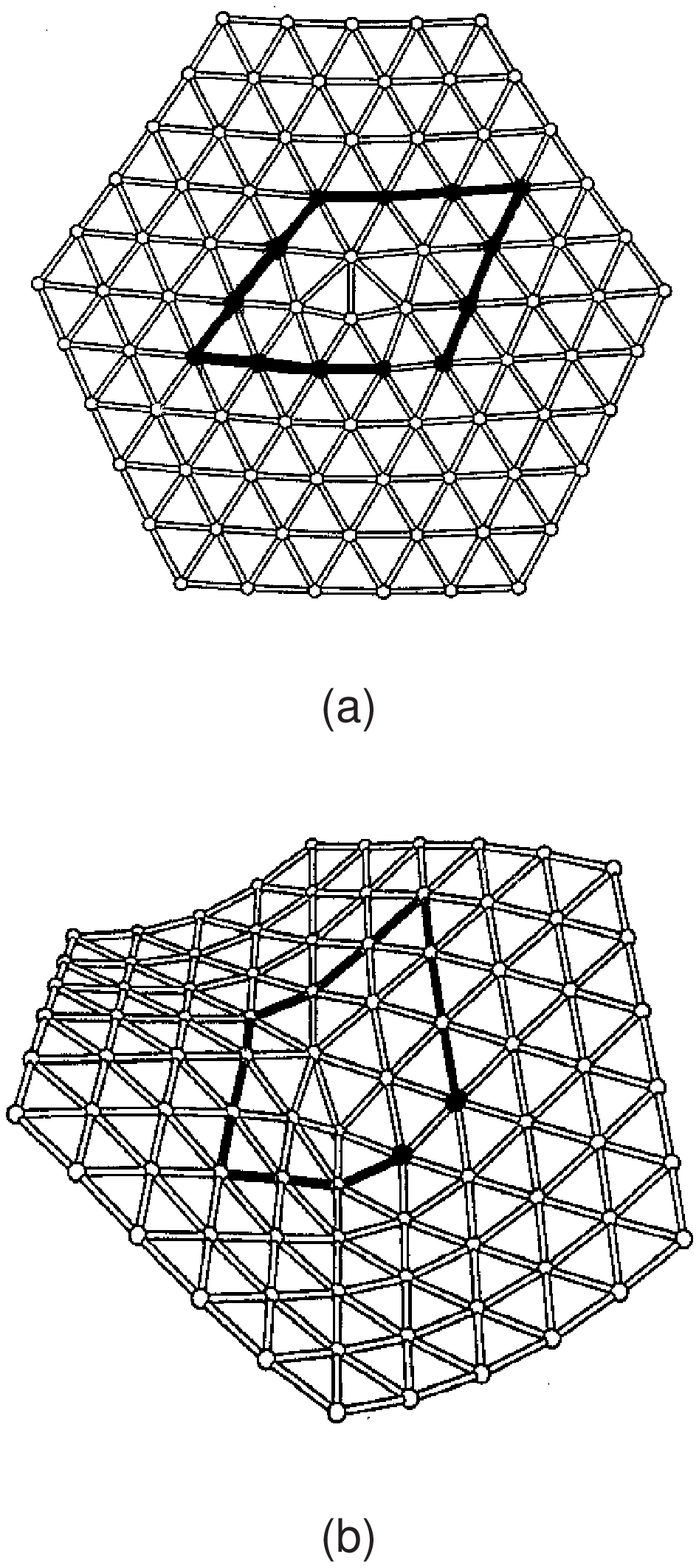}}{FIG.~9. \
Dislocation defect in a (a) monolayer (flat) and a (b) membrane (buckled).
The burgers construction is shown in both cases. From Ref.~[37].}

We can treat dislocations by regarding them as tightly bound disclinations
pairs, or alternatively by starting with the definition of the Burgers vector
in the continuum limit
\begin{eqnarray}
\oint_{\cal C}du_i&=&\oint_{\cal C}
{\partial u_i\over\partial x_j}\; dx_j\nonumber \\
&=&b_i\;,
\eqnum{3.25}
\label{eq:ttwentyfive}
\end{eqnarray}
where (see Fig.~9) the  circuit $\cal C$ encloses a single
dislocation with Burgers vector $\vec b$. For a collection of Burgers
``charges'' $\{\vec b_\alpha\}$ at positions $\{\vr_\alpha\}$, the
differential statement of the noncommutivity embodied in constraints like
(\ref{eq:ttwentyfive}) is
\begin{equation}
\epsilon_{k\ell}\partial_k\partial_\ell u_j=
\sum_\alpha b_{\alpha i}\delta(\vr-\vr_\alpha)\;.
\eqnum{3.26}
\label{eq:ttwentysix}
\end{equation}
Upon noting that the right-hand side of Eq. (\ref{eq:ttwentya}) may be
rewritten,
\begin{equation}
{1\over 2}
\epsilon_{ik}\epsilon_{j\ell}\partial_k
\partial_\ell(\partial_iu_j+\partial_ju_i)=
\epsilon_{k\ell}\partial_k\partial_\ell
\theta+\epsilon_{ip}\epsilon_{k\ell}
\partial_p\partial_\ell\partial_ku_i\;,
\eqnum{3.27}
\label{eq:ttwentyseven}
\end{equation}
we have, combining the above results for disclinations and dislocations,
\begin{equation}
{1\over K_0}\nabla^4\chi(\vr)=
\sum_\alpha\epsilon_{ij}b_{\alpha i}
\partial_j\delta(\vr-\vr_\alpha)+
{\pi\over 3}
\sum_\beta
s_\beta\delta(\vr-\vr_\beta)\;.
\eqnum{3.28}
\label{eq:ttwentyeight}
\end{equation}

Vacancies and interstitials are special cases of a general class of
``impurity defects'' which includes as well substitutional atoms of
the wrong size.  If the defect sits in
a lattice site of 3-fold or higher symmetry, its stress field is
isotropic, and it can be characterized by the area deficit
$\Omega_0$ (positive or negative), it induces in an otherwise
perfect lattice,
\begin{eqnarray}
\Omega_0&=& \int(\vnabla\cdot\vec u)\;d^2r,\nonumber \\
&=&
\oint_{\cal C} \hat n\cdot\vec u(\vr)\;d\ell
\eqnum{3.29}
\label{eq:ttwentynine}
\end{eqnarray}
where $\hat n$ is an outward unit normal to a contour $\cal C$ surrounding
an impurity at the origin. The biharmonic equation for $\chi(r)$
reads \cite{lanlif}
\begin{equation}
\nabla^4\chi(\vr)=-2\mu\Omega_0
\nabla^2\delta(\vr)\;.
\eqnum{3.30}
\label{eq:tthirty}
\end{equation}
A similar equation can be derived for vacancies and interstitials by
regarding them as three dislocations at the vertices of a small
equilateral triangle with radial Burgers vectors pointing
inward (vacancy) or outward (interstitial).

It is straightforward to rewrite the free energy (\ref{eq:televen}) in
terms of $\chi(r)$,
\begin{equation}
F={1\over 2K_0}
\int d^2r[\nabla^2\chi(\vr)]^2\;.
\eqnum{3.31}
\label{eq:tthirtyone}
\end{equation}
Once $\chi(\vr)$ is known, we can thus calculate both the strain
fields (from (\ref{eq:teighteen}) and
the defect energy. For an isolated disclination at the origin with
angle defect $s$, we have \cite{ref37}
\begin{equation}
\chi(\vr)={K_0s\over
8\pi}r^2[\ln(r/a)+{\rm const.}]
\eqnum{3.32}
\label{eq:tthirtytwo}
\end{equation}
and
\begin{equation}
F={K_0s^2\over 32\pi}R^2\;,
\eqnum{3.33}
\label{eq:tthirtythree}
\end{equation}
for a circular patch of crystal with radius $R$. Because disclination strains
do not
fall off at large distances, the free energy diverges {\it quadratically}
with the size of the system. We have neglected a core energy contribution
which is negligible as $R\rightarrow\infty$. For an isolated
{\it dislocation} at the origin, one has
\begin{equation}
\chi(\vr)={K_0\over 4\pi}
b_i\epsilon_{ij}r_j
\ln(r/a)
\eqnum{3.34}
\label{eq:tthirtyfour}
\end{equation}
and
\begin{equation}
F={K_0b^2\over 8\pi}
\ln (R/a) + {\it const.}
\eqnum{3.35}
\label{eq:tthirtyfive}
\end{equation}
The logarithmic divergence arises because the strains fall off like
$1/r$, similar to the superfluid velocity in helium films.

The energy of isolated disclinations is prohibitively large in
crystalline monolayers. This energy is reduced to a logarithmic
divergence in the hexatic monolayers, however, because of screening
by a gas of unbound dislocations \cite{bihdrn78}. The logarithmic
energy of dislocations in two-dimensional crystals leads via the usual
energy/entropy argument to an estimate of the dislocation unbinding
temperature \cite{kosthou72,bere70}. For a more detailed
discussion of the statistical mechanics of defect mediated
melting of monolayers, see Ref.~\cite{drn83}.

For completeness, we note that the Airy stress function of an
isolated impurity at the origin is
\begin{equation}
\chi(\vr)=
{-\mu\Omega_0\over\pi}\ln(r/a)
\eqnum{3.36}
\label{eq:tthirtysix}
\end{equation}
corresponding to a displacement field
\begin{equation}
\vec u(\vr)={\Omega_0\vr\over 2\pi r^2}\;.
\eqnum{3.37}
\label{eq:tthirtyseven}
\end{equation}
Because the strains now fall off as $1/r^2$, the elastic energy is
finite and comparable to the core energy.

\subsection{Defects in Crystalline Membranes}

We now return to the full continuum elastic free energy Eq. (\ref{eq:tnine})
and determine how the defect energies are reduced in membranes, due to
buckling into the third dimension. Variation of $F$ with respect to $f$ and
$\vec u$ leads to the von Karman equations \cite{lanlif}
\begin{equation}
\kappa\nabla^4f={\partial^2\chi\over\partial y^2}
{\partial^2f\over\partial x^2}+
{\partial^2\chi\over\partial x^2}
{\partial^2f\over\partial y^2}-2
{\partial^2\chi\over\partial x\partial y}
{\partial^2f\over\partial x\partial y}
\eqnum{3.38{\rm a}}
\label{eq:tthirtyeighta}
\end{equation}
\begin{equation}
{1\over K_0}\nabla^4\chi=- {\partial^2f\over\partial x^2}
{\partial^2f\over\partial y^2}+
\left({\partial^2f\over\partial x\partial y}\right)^2
+S(\vr)\;.
\eqnum{3.38{\rm b}}
\label{eq:tthirtyeightb}
\end{equation}
The Airy stress function is related to the stress as for monolayers,
$\sigma_{ij}(\vr)=\epsilon_{im}\epsilon_{jn}
\partial_m\partial_n\chi(\vr)$, where we again have
$\sigma_{ij}=2\mu u_{ij}+\lambda u_{kk}\sigma_{ij}$. Now,
however, $u_{ij}$ is the nonlinear stress tensor,
\begin{eqnarray}
u_{ij}&=& {1\over 2}
(\partial_iu_j+\partial_ju_i)+
{1\over 2}{\partial f\over\partial x_i}
{\partial f\over\partial x_j}+
{1\over 2}{\partial \vec u\over\partial x_i}
\cdot{\partial \vec u\over\partial x_j}\nonumber \\
&\approx& {1\over 2}
(\partial_iu_j+\partial_ju_i)+
{1\over 2}{\partial f\over\partial x_i}
{\partial f\over\partial x_j}
\;.
\eqnum{3.39}
\label{eq:tthirtynine}
\end{eqnarray}
In the first line of (\ref{eq:tthirtynine}), we have
included nonlinearities in the in-plane displacements $\vec u$. Although
the nonlinearity in $f$ is {\it always} important for
membranes, the ${1\over 2}{\partial \vec u\over\partial x_i}\cdot
{\partial \vec u\over \partial x_j}$ term can often be neglected.
This terms {\it does} contribute
significantly to disclination energies, however:
In monolayers, for example, it leads to a small correction to the
coefficient of the $R^2$ divergence in the disclination energy
for monolayers \cite{ref37}. In contrast to disclinations, the
strain fields for dislocations and impurities fall off fast
enough to justify neglecting the ${1\over 2} {\partial \vec u\over\partial
x_i}\cdot
{\partial \vec u\over\partial x_j}$ term in evaluating the far-field elastic
energy. We have included a source term on the right-hand side of
Eq. (\ref{eq:tthirtyeightb}) due to defects. For an
isolated defect at the origin, one has
\begin{eqnarray}
S(\vr)=\left\{\matrix{
{\pi\over 3} s\delta(\vr),&{\rm disclination},\nonumber \\
\epsilon_{ij}b_i\partial_j\delta(\vr),&{\rm dislocation}, \nonumber \\
-{2\mu+\lambda\over 2(\mu+\lambda)}\Omega_0\nabla^2
\delta(\vr),&{\rm impurity}, } \right.
\eqnum{3.40}
\label{eq:tforty}
\end{eqnarray}
just as in flat space. The additional contribution to the right-hand side of
Eq. (\ref{eq:tthirtyeightb}) is the Gaussian curvature associated with
the out-of-plane membrane displacements. This ``curvature charge'' can
partially
cancel the ``topological charges'' $S(\vec r)$ of the defects when they
buckle. Note that any solution with $f\not =0$, implies another
solution with $f\rightarrow -f$. When $f=0$, we recover the elastic
equations for monolayers.

\subsubsection{Disclinations and Dislocations }

Equations (3.38) are nonlinear and extremely difficult to solve analytically.
Some progress is possible, however, in the inextensional limit
$K_0\rightarrow\infty$. All defect energies would be infinite in this
case for monolayers. For membranes, however, defects can buckle so as to
eliminate the elastic contributions to Eq. (\ref{eq:tnine}). The only
remaining contribution is the bending energy. The required $f$ field is
determined by setting the right-hand side of Eq. (\ref{eq:tthirtyeightb})
to zero. The inextensional solutions of the von Karman equations representing
a 5-fold disclination are
\begin{equation}
\chi(r)=-\kappa \ln(r/a_0),\qquad
f(r)=\pm\sqrt{1\over 3}\;r\;,
\eqnum{3.41}
\label{eq:tfortyone}
\end{equation}
with energy
\begin{equation}
E_5={1\over 3}\;\pi\kappa\ln(R/a_0)\;,
\eqnum{3.42}
\label{eq:tfortytwo}
\end{equation}
for a circular membrane with radius $R$. A 7-fold  inextensional
disclination can be represented in polar coordinates approximately by
\cite{drnpel,seungdrn}
\begin{equation}
\chi(r)=3\kappa\ln(r/a_0),\qquad
f(\vr)=\pm\sqrt{2\over 9}\; r\sin 2\phi
\eqnum{3.43}
\label{eq:tfortythree}
\end{equation}
which leads to an energy
\begin{equation}
E_7=\pi\kappa\ln(R/a)\;.
\eqnum{3.44}
\label{eq:tfortyfour}
\end{equation}
The $R^2$ divergence of disclination energies in monolayers (with an {\it
infinite}
coefficient when $K_0\rightarrow\infty$!) is thus screened considerably
by buckling.

Precise numerical solutions of the von Karman equations for arbitrary $\kappa$
and $K_0$ can be constructed by minimizing the energy of a
triangulated tethered surface model \cite{seungdrn}. The bending rigidity
is represented by an interaction between neighboring unit normals
to the triangular plaquettes and the in-plane elasticity by nearest neighbor
harmonic springs adjusted to give the correct value of $K_0$. This
``dynamic triangulation'' method is easier to implement than a direct
numerical solution of the von Karman equations, and includes
automatically all relevant nonlinearities. The numerical results for
disclinations in the inextensional limit are similar to Eqs.
(\ref{eq:tfortytwo})
and (\ref{eq:tfortyfour}), with slightly different coefficients
\cite{seungdrn}. It is still true that $F_5<F_7$.
Buckled disclinations obtained using this
approach are shown in Fig.~8. Five-fold disclinations buckle into a
cone, while the seven-fold disclination leads to a saddle surface.
For disclinations and dislocations, the crucial physics lies in the
existence of a buckling radius \cite{drnpel}: For small system radii
$R$, these defects lie flat with no bending. Above a certain critical
radius $R_c$, the defects trade elastic energy for bending energy
and buckle out-of-the-plane. Numerical studies \cite{seungdrn} show
that disclination energies are screened down to a logarithmic
divergence beyond the buckling radius, as suggested by results in the
inextensional limit. Both disclinations and dislocations behave as if
they were inextensional beyond the buckling radius.

Buckling is even more important for dislocations. Beyond the buckling
radius $R_b$, the dislocation energy no longer increases
logarithmically with $R$, and in fact appears to approach a finite
constant \cite{seungdrn}. A buckled dislocation is shown in
Fig.~9. The entropic contribution to the dislocation
free energy $F_d=E_d-S_dT$, however, still
varies logarithmically with system size. The finiteness of the
dislocation energy (or more precisely, {\it any} $R$ dependence of the
energy increasing more slowly than $\ln(R/a)$) then implies that
untethered crystalline membranes must melt
at {\it all} nonzero temperatures \cite{drnpel,seungdrn}.
The areal density of dislocations in a hexatic membrane at temperature
$T$ is approximately
\begin{equation}
n_d\approx a_0^{-2}e^{-E_d/T}
\eqnum{3.45}
\label{eq:tfortyfive}
\end{equation}
where $E_d$ is is the (finite) dislocation energy, and $a_0$ is the
lattice constant. Dislocations in membranes thus behave similarly to
vortices in the ``naive'' model of superconducting films.
Since buckled disclinations still have a logarithmically
diverging energy, the hexatic fluid should be separated from a
high-temperature isotropic liquid by a finite-temperature
disclination unbinding transition (see below). Melting of membranes
confined between two flat plates, which suppresses the buckling
of dislocations and leads to a finite melting temperature $T_m(d)$,
tending  to zero as $d\rightarrow\infty$, has been studied by
Morse and Lubensky \cite{morselub2}.

Defects like dislocations and disclinations buckle into the third
dimension when the cost in stretching energy to remain flat exceeds
the bending energy required to buckle. More generally, buckling
occurs whenever \cite{cardrn}
\begin{equation}
{K_0\ell^2\over\kappa}\ge \gamma
\eqnum{3.46}
\label{eq:tfortysix}
\end{equation}
where $\ell$ is a characteristic length scale for the defect and
$\gamma$ is a dimensionless constant, typically of order $10^2$.
In circular membranes of radius $R$, $\ell=\sqrt{Rb}$ for a
dislocation with Burgers vector $b$, and $\gamma\approx 127$. For
positive disclinations, we have $\ell=R$ and $\gamma\approx 160$, while
for negative disclinations $\ell\approx R$ and $\gamma\approx 192$
\cite{seungdrn}.
Thus isolated dislocations and disclinations {\it always} buckle in
sufficiently large
crystalline membranes, irrespective of the values of the elastic
constants.

Strictly speaking, these conclusions apply only at $T=0$. In a
finite temperature defect-free flat phase, thermal fluctuations
cause the long wavelength wavevector dependent bending rigidity to
diverge, $\kappa_R(q)\sim q^{-\eta_\kappa}$, while the renormalized
elastic parameter $K_R(q)$ tends to zero $K_R(q)\sim q^{\eta_u}$
\cite{drnpel,aronlub,ledous}. It is then appropriate to substitute
$K_0/\kappa \rightarrow K_R/\kappa_R\sim 1/R^{\eta_\kappa+\eta_u}$
in  Eq. (\ref{eq:tfortysix}).
Using the values $\eta_\kappa=0.82$ and $\eta_u=0.36$ \cite{ledous},
we conclude that disclinations $(\ell\sim R)$ will still buckle, while
dislocations $(\ell\sim R^{1/2})$ become asymptotically flat as
$R\rightarrow\infty$. Dislocations could still buckle on short
scales, before these long wavelength  thermal renormalizations of the elastic
parameters
set in. The dislocation energy will in any case tend to a finite
constant for large $R$: Even if the dislocation remains asymptotically
flat, we should now replace $K_0$ by $K_R(R)\sim 1/R^{\eta_u}$ in
Eq. (\ref{eq:tthirtyfive}). Thus the conclusion that the hexatic
phase is the stable low-temperature phase of membranes is unchanged.

\subsubsection{Other Defects in Crystalline Membranes}

Buckling is not inevitable even at $T=0$ for finite energy defects such as
vacancies, interstitials, or impurity atoms. The buckling criterion is
again Eq. (\ref{eq:tfortysix}), where $\ell$ is now related to the
excess area induced by the defects, $\ell\approx
\sqrt{\Omega_0}$ \cite{cardrn}. Buckling can be triggered in an
{\it infinite} system simply by varying the ratio $K_0/\kappa$. Because of the
finite length scale associated with these defects, it is the
bare ``local'' values of $K_0$ and $\kappa$ which should appear in
(\ref{eq:tfortysix}). Since $\kappa$ usually increases with
increasing temperature, and $K_0$ usually decreases, buckling is
most likely at low temperatures. The $f\rightarrow -f$ symmetry of
the von Karman equations insures that buckled defects have at least
two degenerate minima, representing displacements on opposite sides
of the membrane. The ensemble of two-level systems generated by
buckled finite energy defects will contribute to the specific heat
and other equilibrium properties of membranes.

\myfigure{\epsfysize3.5in\epsfbox{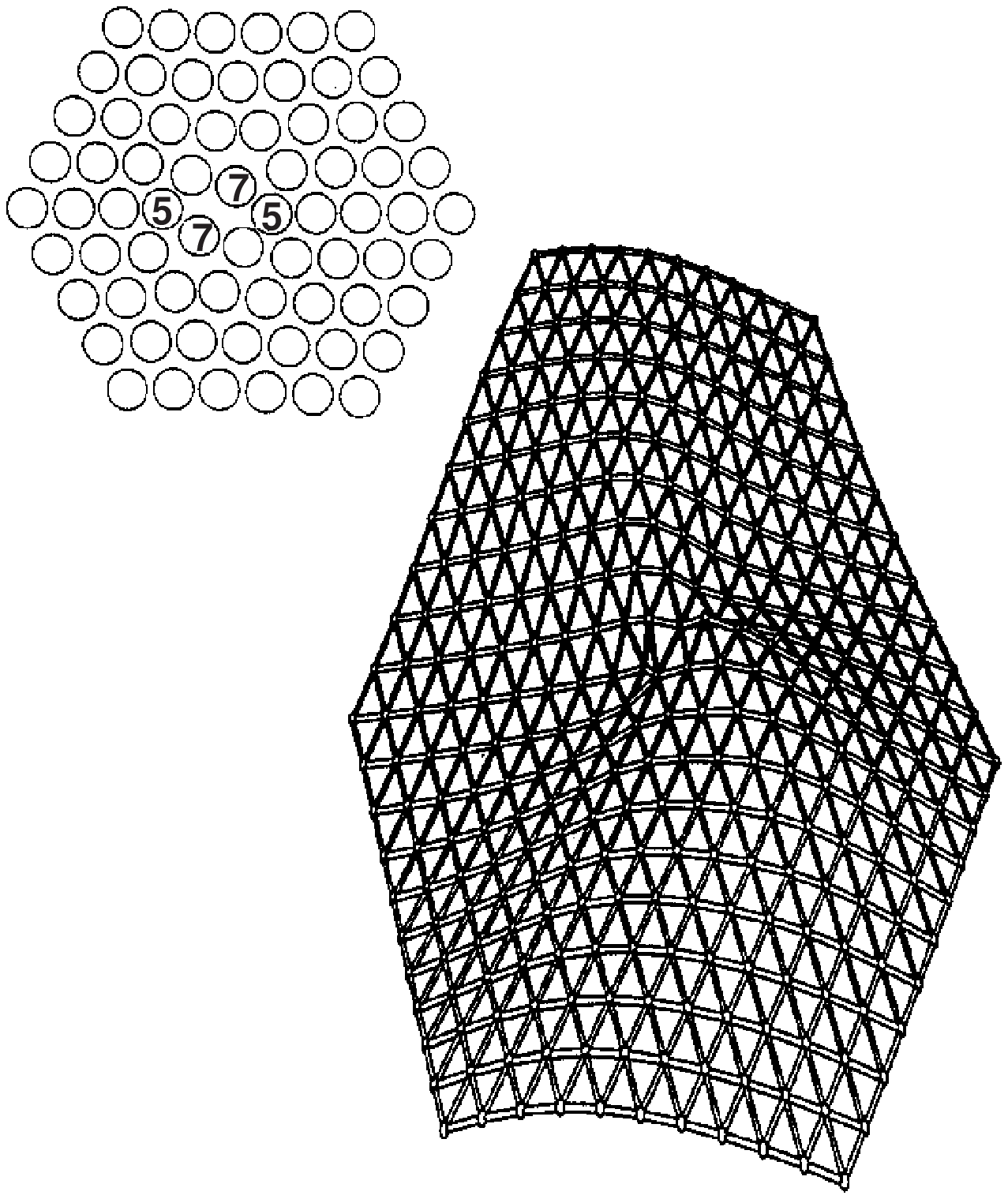}}{FIG.~10. \
Crushed vacancy defect in a flat monolayer and in a membrane.
Note that the initial flat state, consists of two 5's and two 7's,
and can be regarded as a tightly bound dislocation pair. From Ref.~[38].}

The buckling of a vacancy is illustrated in Fig.~10 \cite{cardrn}.
Even in flat space, the vacancy does not have the high symmetry
implied by the  isotropic impurity source term in Eq. (\ref{eq:tforty}). For
the harmonic spring nearest neighbor interaction potential used in
Ref.~\cite{cardrn}, an initially six-fold symmetric vacancy is
``crushed'' into the lower symmetry object indicated in the figure.
The crushed vacancy is equivalent to a tightly bound dislocation
pair. Such vacancies buckle whenever $K_0a_0^2/\kappa \agt 26$
\cite{cardrn}. Figure~10b shows a buckled vacancy for $K_0a_0^2/\kappa=92$.
Note that the 5-fold disclinations have buckled on opposite sides of
the membrane.

It is also interesting to consider buckling of grain boundaries \cite{cardrn}.
Three-dimensional crystals often consist of randomly oriented
grains separated by defect walls. Similar polycrystalline order may
appear in partially polymerized membrane vesicles \cite{cardrn,drnrad,mutz}.
In two dimensions, low-angle grain boundaries can be modeled by a row of
dislocations, with average Burgers vector perpendicular to the
boundary \cite{hirth}. Consider first a grain boundary in a monolayer.
Since all dislocations have the same sign, one might think that the elastic
energy would be enormous. However, cancellations in the long-range
part of the strain field insures a finite elastic energy per unit
length \cite{hirth}. Let $\theta$ be the tilt angle relating the
mismatched crystallites on either side of the grain. A $\theta=21.8^\circ$
grain boundary is shown in Fig.~11a. The spacing $h$ between dislocations
(associated with the large dots in Fig.~11a)
is given by Frank's law \cite{hirth}, $h={1\over 2} b\sin{\theta\over 2}$.
On length scales large compared to $h$, the composite stress field of all the
dislocations dies off exponentially, provided the Burgers vectors
are strictly perpendicular to the boundary; on shorter scales, however, the
behavior of the grain boundary should be dominated by the individual
dislocations.

\myfigure{\epsfysize3.5in\epsfbox{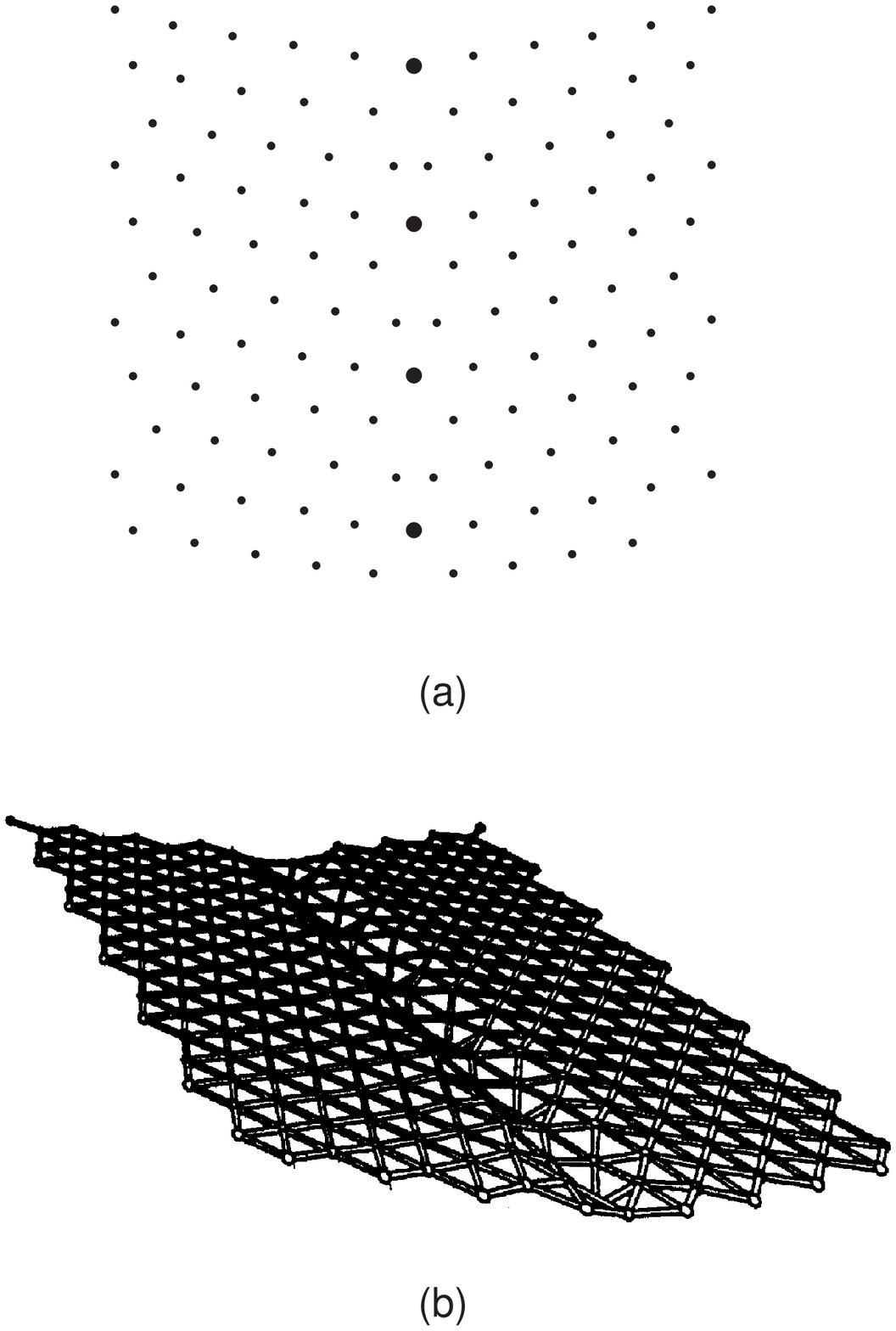}}{FIG.~11. \
Flat (a) and buckled (b) 21.8$^\circ$ grain boundaries. The two
grains meet along the row of heavy dots. From Ref.~[38].}

Consider now a grain boundary inserted into a
{\it membrane}. We have seen that isolated dislocations buckle
whenever $R>R_b\approx 127\kappa/K_0b$. One might expect a similar
buckling in grain boundaries whenever $h> R_b$. Numerical studies
\cite{cardrn} show that Eq. (\ref{eq:tfortysix}) is again satisfied,
with $\ell=\sqrt{bh}$ and $\gamma\approx 120$, consistent with this
guess. A buckled 21.8$^\circ$ grain boundary is shown in Fig.~11b
for $K_0bh/\kappa=300$. Note that the membrane remains {\it flat}
far from the boundary.

\subsection{Defects in Hexatic and Liquid Membranes}

As discussed above, a finite concentration of unbound dislocations is
present in membranes at any nonzero temperature. Provided disclinations remain
bound together, the resulting phase is a hexatic, similar to the hexatic
liquid which arises in the theory of two-dimensional melting of
monolayers \cite{bihdrn78}. Hexatic fluids are characterized by
extended correlations in the complex order parameter $\psi_6(\vr)=\exp
[6i\theta(\vr)]$, where $\theta(\vr)$ is the angle a bond
between neighboring particles centered at $\vr$ makes with respect to a
local reference axis. Because this reference axis changes when
parallel transported on a curved surface, there is an important
coupling between the bond angle field $\theta(\vr)$ and the geometry
of membrane. This coupling stiffens hexatic membranes
relative to their isotropic liquid membrane counterparts, although
they still fluctuate more wildly than a crystalline tethered surface.

The continuum elastic-free energy for hexatic membranes was derived
in Ref. \cite{drnpel}:
\begin{equation}
F_H={1\over 2}K_A\int d^2r
\left[\partial_i\theta-{1\over 2}
\epsilon_{jk}\partial_k((\partial_if)
(\partial_jf))\right]^2+
{1\over 2}\kappa\int d^2r(\nabla^2f)^2
\eqnum{3.47}
\label{eq:tfortyseven}
\end{equation}
where $K_A$ is the hexatic stiffness constant controlling fluctuations in
$\theta(\vr)$. Note the similarity between $F_H$ and the
crystalline membrane energy Eq. (\ref{eq:tnine}). Both free energies contain
a bending rigidity, and the out-of-plane displacements act like a ``vector
potential'' coupled to the low energy Goldstone modes associated with the
relevant broken symmetries. Hexatic membranes have many fascinating
properties, including a ``crinkled'' phase, and an interesting
literature is developing around them [56--58].
Here we discuss the von Karman equations for hexatics and show that they
predict a buckling transition for positive disclination defect at a critical
value of the ratio $\kappa/K_A$ [39,59, 60].
We argue that the energy for negative disclinations must be different, and
point out that this asymmetry has important consequences for liquid
membranes once disclinations proliferate.

The variation of $F_H$ with respect to $f$ and $\theta$ leads to
\begin{equation}
\kappa\nabla^4f=K_A(\partial_i\partial_j f)\epsilon_{jk}
\partial_k(\partial_i\theta-A_i)\;,
\eqnum{3.48}
\label{eq:tfortyeight}
\end{equation}
and
\begin{equation}
\partial_i(\partial_i\theta-A_i)=0,
\eqnum{3.49}
\label{eq:tfortynine}
\end{equation}
with
\begin{equation}
A_i={1\over 2}
\epsilon_{jk}\partial_k((\partial_i
f)(\partial_jf))\;.
\eqnum{3.50}
\label{eq:tfifty}
\end{equation}
In analogy to Eq. (\ref{eq:tsixteen}), we introduce a hexatic stress
function $\chi_H$ (the Cauchy conjugate function for the hexatic ``current'')
via
\begin{equation}
K_A(\partial_i\theta-A_i)\equiv
\epsilon_{ij}\partial_j\chi_H\;,
\eqnum{3.51}
\label{eq:tfiftyone}
\end{equation}
so that Eq. (\ref{eq:tfortynine}) is satisfied automatically. Now apply the
operator $\epsilon_{ik}\partial_{k}$ to Eq. (\ref{eq:tfiftyone}) and
use Eq. (\ref{eq:ttwentytwo}) to rewrite the derivatives of
$\theta$ in terms of the disclination density $S(\vr)$. The resulting
equation, when combined with Eq.~(3.48), results in the ``von
Karman equations for hexatics,''
\cite{seung}
\begin{equation}
\kappa\nabla^4f=
{\partial^2\chi_H\over\partial y^2}
{\partial ^2f\over\partial x^2} +
{\partial^2\chi_H\over\partial x^2}
{\partial^2f\over\partial y^2}-2
{\partial^2\chi_H\over\partial x\partial y}
{\partial^2f\over\partial x\partial y}
\eqnum{3.52{\rm a}}
\label{eq:tfiftytwoa}
\end{equation}
\begin{equation}
-{1\over K_A}\nabla^2\chi_H=
-{\partial^2f\over\partial x^2}
{\partial ^2f\over\partial y^2} +
\left({\partial^2f\over\partial x\partial y}\right)^2
+ S(\vr)\;.
\eqnum{3.52{\rm b}}
\label{eq:tfiftytwob}
\end{equation}
Note that these equations are identical to Eqs. (3.38) except
that $K_0\rightarrow K_A$ and $\nabla^4\rightarrow -\nabla^2$ in
Eq.~(3.52b).

Suppose $S(\vr)=(+\pi/3)\delta(\vr)$, representing a single positive
disclination
at the origin. We then look for functions $\chi_H(\vr)$ and $f(\vr)$ of the
form
\begin{equation}
\chi_H(r)=-\kappa\ln(r/a_0),\qquad
f(r)=\pm\alpha r
\eqnum{3.53}
\label{eq:tfiftythree}
\end{equation}
where $\alpha$ allows for conical dislocation buckling and remains to be
determined. It is easy to check that Eq. (\ref{eq:tfiftytwoa}) is obeyed
for arbitrary $\alpha$. All three terms in Eq. (\ref{eq:tfiftytwob}) are
now proportional to delta functions, and equating the coefficients
determines $\alpha$,
\begin{equation}
\alpha^2={1\over 3}-{2\kappa\over K_A}\;.
\eqnum{3.54}
\label{eq:tfiftyfour}
\end{equation}
For $\kappa/K_A>1/6$, there are no real solutions and the membrane remains
flat. For $\kappa/K_A<1/6$, however, the disclination buckles~\cite{seung}.
Using Eq.~(\ref{eq:tfortyseven}) we readily find  that the energy of this
hexatic
disclination in a membrane of radius $R$ is
\begin{eqnarray}
E_5=\left\{\matrix{
{1\over 3}\pi\kappa(1-3\kappa/K_A)\ln(R/a_0),&\kappa/K_A<{1\over 6},\nonumber
\\
{\pi K_A\over 36}\ln(R/a_0),&\kappa/K_A>{1\over 6}\;.}\right.
\eqnum{3.55}
\label{eq:tfiftyfifty}
\end{eqnarray}
When $\kappa/K_A>1/6$, we recover the elastic energy for disclinations in flat
hexatic monolayers \cite{bihdrn78}. When $K_A\rightarrow\infty$, we recover the
buckled disclination energy (\ref{eq:tfortytwo}) for crystalline membranes in
the
inextensional limit.

\myfigure{\epsfysize5in\epsfbox{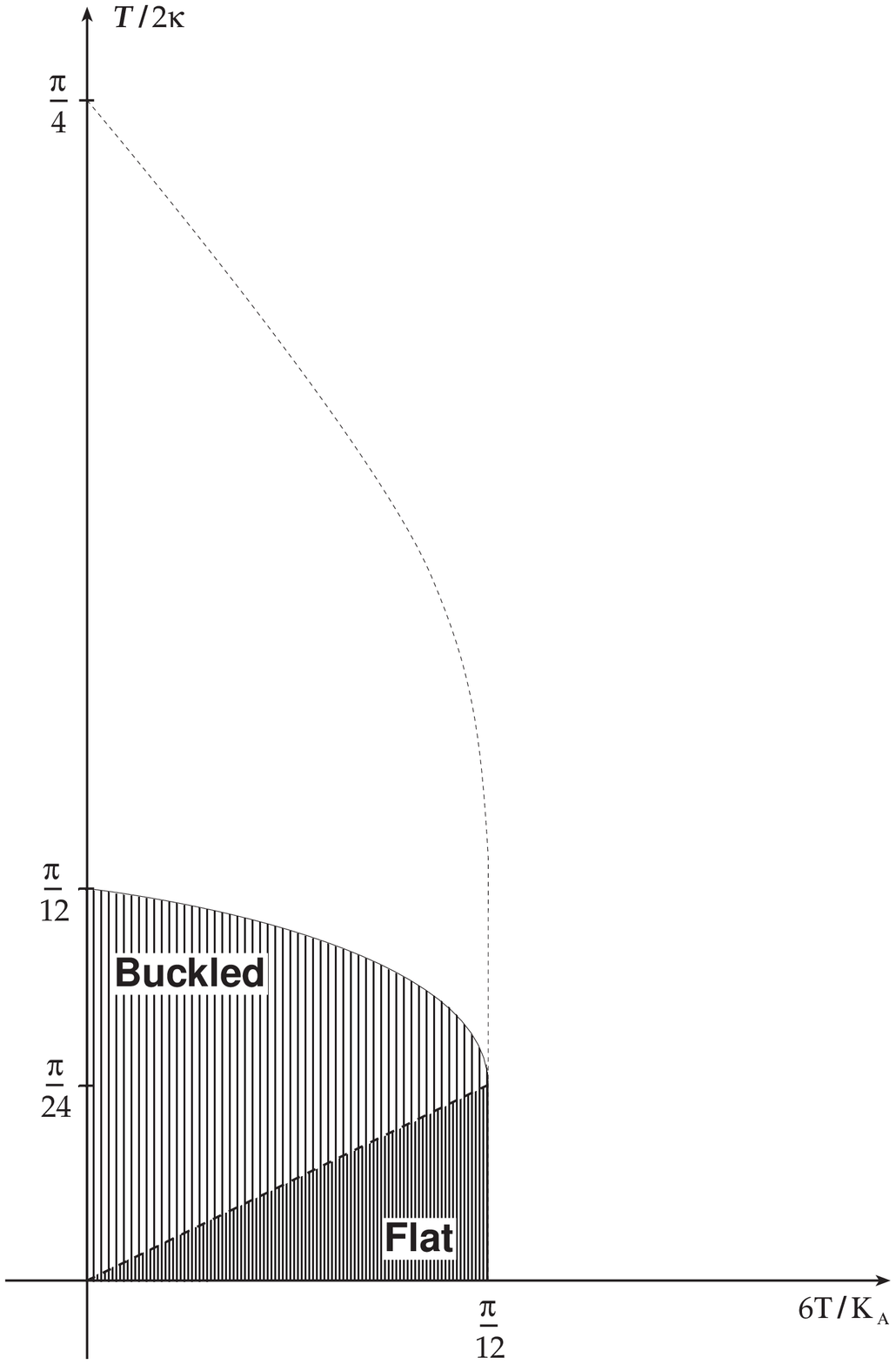}}{FIG.~12. \
Locus of entropic instabilities for a 5-fold disclination in a hexatic membrane
(solid line). The shaded hexatic region is divided into regions where
5-fold disclinations buckle or remain flat, depending on the ratio
$K_A/\kappa$.
The dashed line is an estimate of where an isolated
7-fold disclination becomes
unstable. The region above and to the right of the solid line is an
isotropic fluid.}

The total free energy of the 5-fold disclination including the positional
entropy is
\begin{eqnarray}
F_5&=& E_5-ST\nonumber \\
&=& E_5-2T\ln(R/a_0)\;.
\eqnum{3.56}
\label{eq:tfiftysix}
\end{eqnarray}
This free energy becomes negative above the solid line in Fig.~12. A curve
of this kind was first presented by Guitter and Kardar~[57],
who used a more accurate theory to discuss the consequences of 5-fold
disclination buckling. Their criteria for buckling and for disclination
proliferation are similar, but not identical to ours. Small errors are to
be expected in our approach because the Gaussian curvature $G(r)$ in the
Monge representation is actually
\begin{equation}
G(\vr)={
(\partial_x^2f)(\partial_y^2f)-(\partial_x\partial_yf)^2\over
(1+|\vnabla f|^2)^{1/2}}\;,
\eqnum{3.57}
\label{eq:tfiftyseven}
\end{equation}
in contrast to the small $|\vnabla f|$ approximation which appears in
Eq.~(3.52b) [60].
A factor $(1+|\vec\nabla f|^2)^{1/2}$ should also be included in the
measure~[60].
Nevertheless, the essential predictions of the two approaches are
identical: the hexatic becomes unstable to an entropically driven
proliferation of 5-fold disclinations whenever $\kappa/T$ or $K_A/T$ become
sufficiently small.

What about {\it negative} disclinations? When $K_A\rightarrow\infty$ the
similar
approximations lead
a result identical to the inextensional crystalline 7-fold defect
described by Eq. (\ref{eq:tfortyfour}). Although the energies of {\it both}
5- and 7-fold defects diverge logarithmically, the coefficient for negative
disclinations is {\it larger}. This asymmetry persists (but is reduced
slightly) in more accurate numerical computations \cite{seungdrn}. When
$\kappa\rightarrow\infty$, we approach the monolayer limit, and the
5- and 7-fold disclination energies will diverge logarithmically with
{\it equal} coefficients. It seems plausible that the energy of a 7-fold
disclination diverges logarithmically with system size for {\it arbitrary}
values of $\kappa/K_A$, just as for the 5-fold disclinations. With the
two above limits in mind, it seems clear that the free energy $F_7=
E_7-2T\ln(R/a_0)$ for negative
disclinations only becomes negative above the dashed curve in Fig.~12. The
energy/entropy argument
seems to predict that positive and negative disclinations will
unbind at two distinct temperatures! This conclusion, however, is probably
incorrect.
Consider the region between the solid and dashed lines in Fig.~12. Since
the 5-fold defects are energetically favorable, the positive-free
energy cost of a 7-fold defect can be compensated by a number of nearby
5-fold disclinations: A ``composite defect''
consisting of, say, one 7 and three 5's will have a negative total
free energy. Thus both positive and negative disclinations
will proliferate above the solid line. A more complete theory of the
disclination transition would include interactions between 5's and 7's and
might also  require overall disclination ``charge neutrality.'' Progress in
this direction has recently been made by Park and Lubensky~[58],
although an explicit disclination asymmetry has not yet been incorporated into
the theory.

The fundamental asymmetry between positive and negative disclinations should
persist when hexatics melt into isotropic liquid membranes. Let us denote the
defect core energies by $E_5$ and $E_7$ and assume that all long-range
elastic energies have been screened out. Assume further that there is {\it no}
constraint of disclination charge neutrality. This would be the case in
membranes with free edges, or for the experimentally relevant
case of liquid bilayer surfaces which can change their genus
freely. The areal densities of disclinations will then be different,
\begin{equation}
n_5\approx a_0^{-2}e^{-E_5/T},
\qquad
n_7\approx a_0^{-2}e^{-E_7/T}\;.
\eqnum{3.58}
\label{eq:tfiftyeight}
\end{equation}
Which defect predominates in liquid membranes depends on microscopic details
such as interaction potentials, etc. Close to a transition to  a hexatic phase,
however, the arguments given above suggest that 5-fold disclinations dominate.

Disclination asymmetry in liquid membranes parallels the behavior of vacancies
and interstitials in conventional crystalline solids \cite{ashmer}. Although
vacancies and interstitials are the antidefects of each other, they
nevertheless have very different energies. Unless vacancies and interstitials
are created from a perfect crystal with periodic boundary conditions, they
will in general occur with different concentrations. Periodic boundary
conditions create an artificial constraint which forces vacancies and
interstitials to be created in pairs rather than diffusing in from the
surface. The concentrations must also be equal if the defects are charged, as
in ionic crystals \cite{ashmer}.
The flatness of monolayers similarly constrains the
{\it disclination} densities to be equal at all stages in the theory of
two-dimensional melting \cite{bihdrn78}.
In membranes artificially constrained to have
the topology of a torroidal surface, the numbers of 5- and 7-fold
disclinations are forced to be equal to Euler's theorem. A calculation
similar that for point defects in ionic crystals then leads
to
\begin{equation}
n_5=n_7\approx a_0^{-2}
e^{-(E_5+E_7)/2T}\;.
\eqnum{3.59}
\label{eq:tfiftynine}
\end{equation}

The free energy of liquid membranes is often expressed in the Helfrich
form~[62],
\begin{equation}
F_L={1\over 2}
\kappa\int d^2r(\nabla^2f)^2+
\kappa_G\int d^2r
[(\partial_x^2f)
(\partial_y^2f)-(\partial_x\partial_yf)^2]\;.
\eqnum{3.60}
\label{eq:tsixty}
\end{equation}
The first term is the usual bending rigidity. The remaining one is
proportional to the integrated Gaussian curvature and its coefficient
$\kappa_G$
is often called the Gaussian rigidity. This second term is a perfect
derivative, and integrates to a constant for surfaces of fixed genus
\cite{drnwein}. The microscopic origins of $\kappa_G$ are
obscure. Its sign, however, clearly determined by  the disclination asymmetry
discussed
above: An excess of 5-fold disclinations corresponds to $\kappa_G<0$ and favors
{\it positive} net Gaussian curvature. Membrane phases many spherical
vesicles will predominate in this case. An excess of 7-fold disclinations
means $\kappa_G>0$, and a bias toward negative Gaussian curvatures.
Complex ``plumbers nightmare'' lipid membrane phases \cite{drnwein} are then
favored.

On a more formal level, it must be the case from Euler's theorem that the
Gaussian
curvature integrated over a membrane with free edges of area $\Omega$ gives the
disclination asymmetry
\begin{equation}
\int d^2r\sqrt{g} G(\vec r) =
\Omega(n_5-n_7)\;,
\eqnum{3.61}
\label{eq:tsixtyone}
\end{equation}
where $\sqrt{g}=1+|\vec\nabla f|^2$ and $G(\vec r)$ is given by
Eq. (\ref{eq:tfiftyseven}). Equation~(\ref{eq:tsixty}) can thus be rewritten as
\begin{equation}
F_L={1\over 2}\kappa\int d^2r
(\nabla^2f)^2+\kappa_G(N_5-N_7)
\eqnum{3.62}
\label{eq:tsixtytwo}
\end{equation}
where $N_5$ and $N_7$ are the total numbers of 5- and
7-fold disclinations in the membrane. The Gaussian rigidity $\kappa_G$
thus acts as a chemical potential which must be adjusted to give the
correct asymmetry between the populations of 5's and 7's. In this sense, its
effect is similar to that of a nonzero magnetic field in
Eq.~(\ref{eq:fiftysix}),
which would lead to a net excess of positive or negative vortices in
superconducting films. For a related perspective on the physics of metallic
glasses, see Ref.~\cite{drnspap}.

It would be interesting to search for the disclination asymmetry discussed
here in computer simulations of membranes. The most straightforward
approach would be to study initially flat (i.e., large $\kappa$) liquid
membranes with free edges, so that disclinations can enter and exit
freely at the boundary. When the bending rigidity is reduced, a bias
in the average Gaussian curvature
should emerge as the membranes curl up into the third dimension. One could then
vary the interparticle  potentials to study what factors influence the sign of
$E_7-E_5$.
Such an understanding might lead to controlled synthesis of membranes with a
predetermined
sign of $\kappa$.

\acknowledgements

I am grateful for the advice of E. Guitter, M. Kardar, T.C. Lubensky and
J.D. Reppy while preparing this review. This work was supported by the
National Science Foundation, through Grant No. DMR--9417047, and in part
through the Harvard Materials Research Science and Engineering Center via
Grant DMR--9400396.
\bigskip
\bigskip
\newpage
\centerline{\bf APPENDIX: SUPERFLUID DENSITY AND MOMENTUM CORRELATIONS}
\vskip0.5truein
\def\hch{{\hat {\cal H}}}
\def\vu{{\vec u }}
\def\vp{{\vec P }}
\def\vg{{\vec g }}
\def\vr{{\vec r }}
\def\vq{{\vec q }}

\myfigure{\epsfysize3.5in\epsfbox{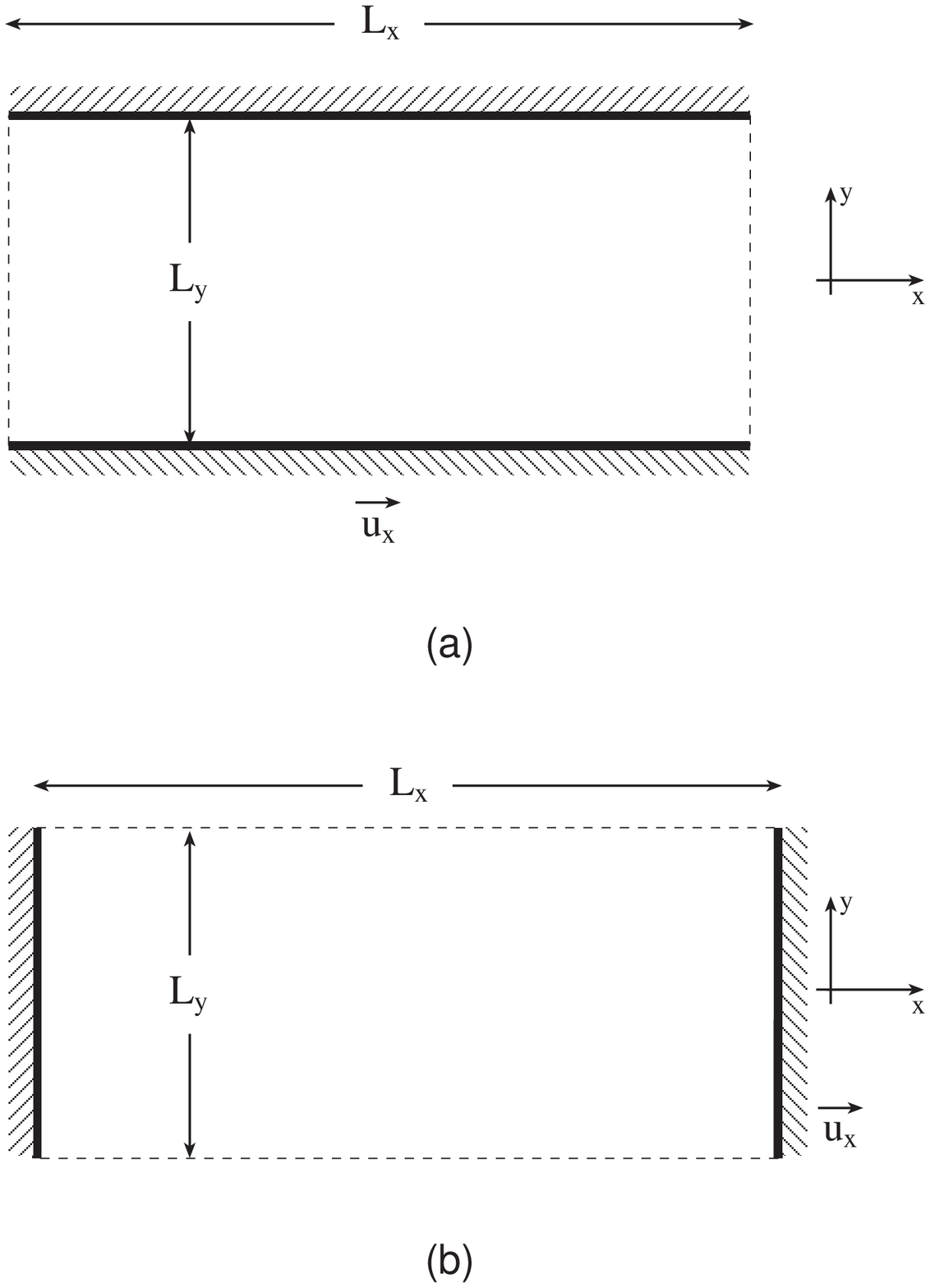}}{FIG.~13. \
Two experiments which lead to the superfluid density.
In (a), there are periodic boundary conditions in the
$x$ direction, and impenetrable walls at $y=\pm L_y/2$. These boundary
conditions are reversed in (b). In both cases, the walls and substrate move at
velocity $u_x$ in the $x$ direction.}

Consider a simplified version of the oscillating substrate experiment
discussed in Sec.~IIA.
Imagine that the substrate is wrapped around to form a cylinder, coated
uniformly with He$^4$. We neglect substrate inhomogeneities, as is appropriate
near the transition or if the films are sufficiently thick~[63].
Now imagine that the substrate oscillations are around
the cylinder axis, and slow enough so that the motion of the substrate is
essentially a uniform translation with velocity $\vu$. As illustrated in
Fig.~13a,
we choose a coordinate system such that the substrate moves along the
$x$ axis. The cylinder has height $L_y$ along the $y$ axis,
and there are barriers
at $y=\pm L_y/2$ which prevent the film from escaping in this direction.
Its circumference is $L_x$.
In two {\it or} three dimensions, the superfluid density measures the
response of liquids to moving walls \cite{hoh,baym,fos}.
In Fig.~13a,
the ``walls'' are provided by the substrate itself plus the barriers at the
top and bottom of the cylinder. Instead of solving a complicated statistical
mechanics problem with moving boundaries, it is easier to make a Galilean
transformation to a coordinate system which moves at the velocity of the
substrate, so that the boundaries are fixed. In the original laboratory frame
of reference, the superfluid fraction of the film will remain at rest. Hence,
it appears to be moving with average velocity $-\vec u$ after the Galilean
transformation.

If $\hch$ is the Hamiltonian in laboratory frame, averages after the
Galilean transformation must be computed with respect to the Hamiltonian~[64].
\begin{equation}
\hch '=\hch -\vu \cdot \vp +{1\over2}\; Mu^2\; ,
\eqnum{A1}
\label{eq:A1}
\end{equation}
where $\vp$ is the total momentum of the film and $M$ is the total mass. Upon
coarse graining the film, the free energy which appears in Eq.~(\ref{eq:ten})
is correspondingly replaced by
\begin{equation}
F'=F-\vu \cdot \int \vg (\vr )d^2r\;+O(u^2)\; ,
\eqnum{A2}
\label{eq:A2}
\end{equation}
where $\vg(\vr)$ is given by Eq.~(2.43).
The relative weights of different configurations of $\psi (\vr )$ in the
moving frame with stationary boundary conditions, given by $\exp [-F'/T]$,
must be the same as those in the laboratory frame with moving boundary
conditions. The average value of the momentum in the laboratory frame is
thus given by
\begin{equation}
\langle g_i(\vr)\rangle_{\vu}={\int {\cal D} \psi (\vr )g_i(\vr)\
e^{-F'/T}\over
\int {\cal D} \psi (\vr )\ e^{-F'/T}}\; ,
\eqnum{A3}
\label{eq:A3}
\end{equation}
where $\vg =(\hbar/2i)[\psi^*(\vr )\vec
\nabla \psi (\vr )-\psi (\vr )\vec
\nabla \psi (\vr )]$ and the subscript on the average denotes a system with
boundaries moving at velocity $\vu$. To linear order in the wall velocity,
we then have
\begin{equation}
\langle g_i(\vr)\rangle_{\vu}={1\over T}\ \int d^2r'\;
\langle g_i(\vr )g_j(\vr \;')
\rangle_{\vec u=0}\ u_j+O(u^2)\; ,
\eqnum{A4}
\label{eq:A4}
\end{equation}
and we see that the momemtum generated by moving walls in a helium film is
determined by the momentum correlations for a system with the walls at rest.
We now expand $\vg (r)$ in Fourier variables,
\begin{equation}
\vg (\vr )={1\over \Omega_0}\sum_{\vq}g(\vq )e^{i\vq \cdot \vr}\;,
\eqnum{A5}
\label{eq:A5}
\end{equation}
with
\begin{equation}
g(\vq )=\int d^2r\ e^{-i\vq \cdot \vr}g(\vr )\;,
\eqnum{A6}
\label{eq:A6}
\end{equation}
and where $\Omega_0$ is the system area. Equation~(\ref{eq:A4})
can now be rewritten as
\begin{equation}
\langle g_i(\vr )\rangle_{\vu}={1\over T}\;\lim_{\vq \rightarrow 0}\;
C_{ij}(\vq )u_j+O(u^2)
\eqnum{A7}
\label{eq:A7}
\end{equation}
where
\begin{eqnarray}
C_{ij}(\vq )&=&\langle g_i(\vq )g_j(-\vq )\rangle_{\vu =0}\nonumber\\
&\equiv& A(q){q_iq_j\over q^2}+B(q)\left(\delta_{ij} -{q_iq_j\over
q^2}\right)\; ,
\eqnum{A8}
\label{eq:A8}
\end{eqnarray}
and $A(q)$ and $B(q)$ are functions only of the {\it magnitude} of $\vq$.

The coefficient of $u_j$ Eq.~(\ref{eq:A7}) involves a delicate limiting
procedure whenever $A(q)\ne$ $B(q)$ \cite{hoh,baym,fos}.
Consider first the situation outlined above, with the periodic boundary
conditions in the x-direction appropriate to a smooth cylindrical substrate.
For any finite cylinder circumference $L_x$, there are always Fourier
components of the $\vg (\vr )$ at $q_x=0$, projected out by the $x'$
integration in Eq.~(\ref{eq:A4}). Along the $y$ direction, however, the $q_y=0$
mode only appears in the limit $L_y\to \infty$, because $\vg (\vr )$ must
vanish at the top and bottom of the cylinder. Thus, the correct order of
limits for large sample sizes in this experiment is the limit $q_x\to 0$,
followed by the limit $q_y\to 0$. The $x$ component of the film momentum
induced by the moving substrate is evidently
\begin{eqnarray}
\langle g_x(\vr )\rangle_{\vec u}&=&{1\over T}\;
\lim_{q_y\to 0}\;\lim_{q_x\to 0}\;C_{xx}
(\vq )u_x\nonumber \\
&=&{B(0)\over T}\,u_x\; .
\eqnum{A9}
\label{eq:A9}
\end{eqnarray}
Since only the normal component of the film moves with the substrate, we
identify $\rho_n(T)$ with the coefficient of $u_x$,
\begin{equation}
\rho_n(T)={B(0)\over T}\; .
\eqnum{A10}
\label{eq:A10}
\end{equation}

Now consider a related, but fundamentally different experiment: We repeat
the cylinder periodically along the $y$ axis, but erect  impenetrable
barriers along the $x$ axis at  fixed positions $x=\pm L_x/2$. See
Fig.~13b. The new barrier
(equivalent to gouging a slit in the substrate parallel to the axis of the
cylinder) prevents the film from circulating completely around the cylinder.
The correct limiting procedure is now the limit $q_y\to 0$, reflecting the
periodic boundary conditions along $y$, followed by the limit $q_x\to 0$,
describing the new barriers in the limit $L_x\to \infty$,
\begin{eqnarray}
\langle g_x(\vr )\rangle&=&{1\over T}\;\lim_{q_x\to 0}\;\lim_{q_y\to 0}\;
C_{xx}(q)u_x\nonumber\\
&=&{A(0)\over T}\,u_x\; .
\eqnum{A11}
\label{eq:A11}
\end{eqnarray}
In this experiment, {\it all} of the film must clearly move at the substrate
velocity since it is pushed around by the barrier. Now we identify
$\rho_{\rm tot}$, the total film density, with the coefficient of $u_x$,
so that
\begin{equation}
\rho_{\rm tot}={A(0)\over T}\; .
\eqnum{A12}
\label{eq:A12}
\end{equation}

In a normal liquid, these two different limiting procedures would lead to
identical results. In a superfluid, however, there is a nonzero superfluid
density, defined by
\begin{equation}
\rho_s(T)=\rho_{\rm tot}-\rho_n(T)\; .
\eqnum{A13}
\label{eq:A13}
\end{equation}
The formula for the superfluid density is thus
\begin{equation}
\rho_s(T)={1\over T}\;\lim_{q\to 0}\;[A(q)-B(q)]\; .
\eqnum{A14}
\label{eq:A14}
\end{equation}


\begin{references}
\bibitem{eq2-12} This similarity becomes clearer if one rewrites Eq.~(2.12)
as $\psi(\vec r)\approx \psi_0[1+i\theta(\vec r)]$ for small $\theta (r)$.
\bibitem{substrate} Substrate disorder introduces  quenched random fluctuations
into the coefficient of $|\psi|^2$ in Eq.~(2.11). Such randomness is known to
be irrelevant if the specific heat exponent $\alpha$ is negative. The essential
singularity in the specific heat at $T_c$ may be interpreted as
$\alpha=-\infty$.
\end{references}
\end{document}